\documentclass[twocolumn,trackchanges]{aastex701}
\usepackage{graphicx}
\usepackage{amsmath}
\usepackage{natbib}
\usepackage{amssymb}
\usepackage{subfigure}
\usepackage{subcaption}
\usepackage{hyperref}

\shorttitle{Jets from proto-magnetars}
\shortauthors{D.~K.~Desai, L.~Combi, D.~M.~Siegel, \& B.~D.~Metzger}

\begin{document}

\newcommand{\be}{\begin{equation}}
\newcommand{\ee}{\end{equation}}

\newcommand  \gcc {~\mathrm{g}~\mathrm{cm}^{-3}}
\newcommand  \E {\times 10^}
\newcommand  \ms {~\mathrm{ms}}
\newcommand  \mdot {~M_\odot~\mathrm{s}^{-1}}

\title{Relativistic jets from millisecond proto-magnetars}

\author[0000-0002-8914-4259]{Dhruv K.~Desai}
\email{desaidhruv5@gmail.com}
\affil{Institute of Physics, University of Greifswald, D-17489 Greifswald, Germany}
\author[0000-0002-5427-1207]{Luciano Combi}
\email{combi.luciano@gmail.com}
\affiliation{Perimeter Institute for Theoretical Physics, Waterloo, Ontario N2L 2Y5, Canada}
\affiliation{Department of Physics, University of Guelph, Guelph, Ontario N1G 2W1, Canada}
\author[0000-0001-6374-6465]{Daniel M.~Siegel}
\email{daniel.siegel@uni-greifswald.de}
\affiliation{Institute of Physics, University of Greifswald, D-17489 Greifswald, Germany}
\affiliation{Department of Physics, University of Guelph, Guelph, Ontario N1G 2W1, Canada}
\author[0000-0002-4670-7509]{Brian D.~Metzger}
\email{bdm2129@columbia.edu}
\affil{Department of Physics and Columbia Astrophysics Laboratory, Columbia University, Pupin Hall, New York, NY 10027, USA}
\affil{Center for Computational Astrophysics, Flatiron Institute, 162 5th Ave, New York, NY 10010, USA}

\begin{abstract}
Rapidly rotating, strongly magnetized neutron stars (``millisecond proto-magnetars'') formed in stellar core-collapse, neutron star mergers, and white dwarf accretion-induced collapse have long been proposed as central engines of gamma-ray bursts (GRB) and accompanying supernovae/kilonovae. However, during the first few seconds after birth, neutrino heating drives baryon-rich winds from the neutron star surface, potentially limiting the magnetization and achievable Lorentz factors of the outflow and casting doubt on whether proto-magnetars can launch ultra-relativistic jets at early times, as needed to power short-duration GRB.  We present three-dimensional general-relativistic magnetohydrodynamic simulations of neutrino-heated proto-magnetar winds that incorporate M0 neutrino transport. While the global wind properties broadly agree with previous analytic estimates calibrated to one-dimensional models, our simulations reveal essential multidimensional effects. For rapidly rotating models with spin periods $P \approx 1\,\mathrm{ms}$, centrifugal forces strongly enhance mass loss near the rotational equator, producing a dense, sub-relativistic outflow ($v \sim 0.1c$). This equatorial wind naturally confines and collimates less baryon-loaded outflows emerging from higher latitudes, leading to the formation of a structured bipolar jet with a peak magnetization along the pole up to $\sigma \sim 30-100$, sufficient to reach bulk Lorentz factors $\Gamma_{\infty} \sim 100$ on larger scales.  The resulting angular stratification of the outflow energy into ultra-relativistic polar and sub-relativistic equatorial components is broadly consistent with the observed partition between beaming-corrected GRB energies and supernova/kilonova ejecta. Our results demonstrate that millisecond proto-magnetars can launch relativistic jets within seconds of formation and highlight their potential role in powering the diverse electromagnetic counterparts of compact-object explosions.
\end{abstract}

\section{Introduction}

Proto-neutron stars (PNSs) are the hot, dense remnants created in the aftermath of the core-collapse of massive stars (e.g., \citealt{Burrows&Lattimer86,Pons+99}), the accretion-induced collapse (AIC) of white dwarfs (e.g., \citealt{Woosley&Baron92}), and from the mergers of binary neutron stars (e.g., \citealt{Dessart+09}). They can contain up to $10^{52}-10^{53}$ erg of rotational energy, for spin periods $P \sim 1$ ms close to the centrifugal break-up limit. Dynamo processes may tap this energy to amplify large-scale magnetic fields to values $\gtrsim 10^{15}$ G \citep{duncan_formation_1992,thompson_neutron_1993,white_origin_bfield_2022}. Recent global simulations of mergers and AIC reveal the formation of an ordered magnetic field of this magnitude due to an alpha-omega dynamo driven by the magneto-rotational instability \citep{combi_rprocess_2023,combi_jet_2023,kiuchi_dynamo_2024,combi_aic_2025}. Such ``millisecond proto-magnetars'' have been variously linked to $r$-process heavy element production, both during the early neutrino-wind phase \citep{thompson_magnetic_2003,prasanna_r_process_2025} and in giant flares at much later stages in their evolution \citep{Cehula+24,patel_direct_evidence_2025}, and even to heavy cosmic ray production \citep{Arons03,Metzger+11b,farrar_uhecr_2025,Patel+26}.

Depending on their rotation rates and magnetic field strengths, millisecond proto-magnetars have also been invoked as power sources behind superluminous supernovae (SLSNe; \citealt{Kasen&Bildsten10,Woosley10}) and gamma-ray bursts (GRB; \citealt{usov_millisecond_1992, wheeler_asymmetric_2000,Metzger+11a}).  In GRBs, the seconds-long gamma-ray emission is powered by dissipation of energy within a collimated relativistic jet from a compact object central engine (e.g., \citealt{Paczynski_winds_1990, Beloborodov10}). Millisecond magnetars are promising engines of GRBs \citep{komissarov_magnetar-energized_2007}, as they can in principle be formed during the (likely magnetically driven; \citealt{LeBlanc&Wilson70, meier_magnetohydrodynamic_1976, wheeler_asymmetric_2002,burrows_simulations_2007,Mosta+14}) explosion of massive stars (e.g., \citealt{Woosley&Bloom06}), as the remnants of neutron star mergers \citep{lee_progenitors_2007, bucciantini_short_2012}, and, potentially, during the AIC of rapidly spinning white dwarfs \citep{usov_millisecond_1992,Blackman&Yi98,Metzger+08a,Cheong+25,combi_aic_2025}. 

Historically, GRBs have been classified observationally into two main categories, short ($\lesssim 2$ s) and long duration ($\gtrsim 2$ s) \citep{kouveliotou_identification_1993}, of likely distinct physical origin. Much evidence supports at least some short GRBs originating from binary neutron star mergers \citep{berger_environments_2011} with associated kilonova (KN) emission \citep{li_transient_1998,metzger_electromagnetic_2010}, while long GRBs are associated with core-collapse of stripped massive stars and associated supernovae \citep{Woosley&Bloom06}.  Recently, the dichotomy of long and short GRBs has been blurred, as some “long” GRBs were followed by KN-like counterparts \citep{Rastinejad+22,Troja+22,Levan+24}.  Even decades ago it was appreciated that short GRBs are sometimes followed by variable X-ray/gamma-ray emission lasting around a minute after the initial burst \citep{gal-yam_novel_2006,gehrels_new_2006,levan_long_short_divide_2007,perley_grb_2009}.  While a millisecond magnetar \citep{Metzger+08a} or accreting black hole \citep{Gottlieb+23} might be capable of powering this long-lived jetted emission\footnote{For example, \citet{desai_imprints_2019} suggest that $r-$process heating of weakly bound tidal ejecta could extend the timescale over which mass falls back to the black hole after a merger, causing a delay in the accretion powered X-rays.  \citet{ciolfi_short_2015} outline a ``time-reversal'' scenario in which a leptonic pulsar wind driven by a proto-magnetar irradiates SN ejecta over long timescales but is initially trapped, allowing the radiation to reach the observer even after a black hole (responsible for the initial GRB) has formed (see also \citealt{siegel_electromagnetic_2016-1,siegel_electromagnetic_2016-2})}, this raises the question of what powers the distinct initial ``short'' phase of the GRB prompt emission lasting a few seconds or less \citep{ciolfi_short_2015,Gottlieb+25b}. 

In order to power a short GRB from a merger or AIC event, a young millisecond magnetar must be capable of placing a significant fraction of its spin-down power into an ultra-relativistic collimated outflow (``jet'') during the first few seconds after forming.  One challenge is that at such early times, the PNS is hot and emits a strong flux of neutrinos \citep{Dessart+09,Roberts&Reddy17}, which heat matter above the neutron star surface and load any outflow with baryons (the so-called ``neutrino wind''; \citealt{Qian&Woosley96}).  Magnetic fields and millisecond rotation greatly boost the power of the magnetar outflow relative to a weakly magnetized PNS. Nevertheless, this neutrino-driven mass-loading of the open field lines limits the outflow's magnetization, $\sigma$, and corresponding maximum achievable Lorentz factor, $\Gamma_{\infty} \le \sigma$.  Compactness arguments applied to on-axis (cosmological) GRBs place lower limits on the jet Lorentz factors $\Gamma \gtrsim 10-100$ (e.g., \citealt{piran_gamma-ray_1999,Lithwick&Sari01,nakar_short-hard_2007}), while for the off-axis merger GW170817, a lower limit of $\Gamma \gtrsim 4$ on the jet was required based on VLBI imaging of the radio afterglow (e.g., \citealt{Mooley+18}).  

Semi-analytic magnetar wind models, calibrated to one-dimensional simulations \citep{thompson_magnetar_2004,metzger_proto-neutron_2007}, predict $\sigma \lesssim 1$ during the earliest stages after millisecond magnetar birth (\citealt{Metzger+11a}; their Fig.~2). 
However, such models treat the wind through its angle-averaged properties, and assume a split monopole outflow geometry. In fact, centrifugal effects in the neutrino heating region endow the most rapidly spinning magnetar winds with a strong latitudinal dependence (e.g., \citealt{Metzger+08}).  Previous axisymmetric magnetohydrodynamic (MHD) simulations \citep{Bucciantini+06,komissarov_magnetic_2007,Prasanna+23,prasanna_r_process_2025} capture some features of this angular dependence, and self-consistently determine the open/closed field topology of the magnetosphere (and its potential time-dependence; \citealt{thompson_magnetic_2003,Prasanna+22,desai_mag_2023}).  However, previous works either employ simplified inner boundary conditions (e.g., fixing the inner temperature of the wind, instead of accounting self-consistently for neutrino heating and cooling), or they neglect special-relativistic effects, precluding the formation of relativistic outflows.  

Past studies have also addressed the interaction of the magnetar wind with the expanding supernova or merger ejecta on much larger spatial scales (e.g., \citealt{Uzdensky&MacFadyen07, Bucciantini+07,Bucciantini+08,Bucciantini+09,bucciantini_short_2012}), to explore whether the external environment can collimate the magnetar wind into a bipolar jet.  However, three-dimensional MHD instabilities, which act beyond the wind termination shock, call into question the effectiveness of jet formation and collimation of a quasi-spherical magnetar wind (e.g., \citealt{porth_three-dimensional_2014,Mosta+14}).  As we find here, collimation by the external environment is not needed for the fastest spinning magnetar engines: the neutrino-wind itself, particularly slow, heavily mass-loaded outflows from lower latitudes, help confine and direct the polar magnetic field lines along the spin axis.  Aside from their role in shaping the jet, the equatorially-concentrated magnetar wind is powerful and appreciably contributes to the total energetics of the surrounding explosion and the quantity of neutron-rich ejecta capable of powering kilonova emission \citep{thompson_magnetar_2004,metzger_proto-neutron_2007,metzger_magnetar_2018,combi_jet_2023}.

In this Letter, we present the first three-dimensional general-relativistic MHD (GRMHD) simulations of the outflows from millisecond proto-magnetars in order to address the key question of whether these objects can produce relativistic jets within the first few seconds after their formation.  Building on our previous work on rapidly spinning but weakly magnetized (\citealt{desai_pnsw_2022}, henceforth Paper I) and strongly magnetized but slowly rotating (\citealt{desai_mag_2023}, henceforth Paper II) PNS winds, we explore the combined impact of rapid rotation and strong magnetic fields on the steady-state wind properties across a range of surface magnetic field strengths and rotation rates. Section \ref{sec:methods} summarizes the computational setup and initial conditions. Section \ref{sec:results} presents results from a suite of models, focusing on the processes of relativistic jet formation in our most highly magnetized and rapidly spinning models.  A more in-depth study of the wind properties, including their nucleosynthetic yields, will be provided in a follow-up paper (Desai et al., in prep). Section \ref{sec:discussion} presents the implications of our results to GRBs, and in Section \ref{sec:conclusions} we deliver our conclusions.


\section{Numerical Methods and Simulation Suite}
\label{sec:methods}

We evolve the GRMHD equations in flux-conservative form with a modified version of the open-source \texttt{GRHydro} code \citep{mosta_grhydro_2014}, which is part of the \texttt{Einstein Toolkit} \citep{loffler_einstein_2012}. Our code \citep{siegel_three-dimensional_2018,combi_rprocess_2023,combi_aic_2025} solves the ideal GRMHD equations with a finite-volume scheme using WENO-Z reconstruction and the approximate HLLE Riemann solver for the hydrodynamic variables by default. The magnetic field is evolved through the electromagnetic 4-potential in the generalized Lorenz gauge \citep{farris_binary_2012} using an upwind constraint transport scheme \citep{del_zanna_echo_2007} for the electric field. The 4-potential is discretized on staggered grids relative to the hydrodynamic variables \citep{etienne_illinoisgrmhd_2015}. This allows us to maintain the solenoidal constraint to machine precision, even across refinement level boundaries and over arbitrarily long timescales; it represents a significant improvement to the constrained transport scheme used in Papers I and II, where accumulating constraint violations across refinement level boundaries complicated long-term evolution. 
Primitive variables are recovered from conservative ones with the framework presented in \citet{siegel_recovery_2018}, which supports finite-temperature, composition-dependent equations of state (EOS). We use the SFHo nuclear EOS \citep{Steiner+13} in tabulated form \citep{schneider_open-source_2017}, which we extend to very small rest-mass densities, $\rho \gtrsim 10^{-2} \gcc$, using the Helmholtz EOS \citep{timmes_accuracy_2000} following the approach in \citet{hayashi_seconds_grmhd_2022}.

To maintain numerical stability in the presence of fast, highly magnetized outflows, we use radially dependent density floors of the form $\rho = \rho_o(r/{9 ~\rm km})^m$, with parameters $\rho_o$ and $-4 \le m \le -2$ chosen for each model to ensure that the atmosphere remains at least an order of magnitude below the wind profile. We find our results unaffected by a variety of density floors tested, as the floor is not reached in the computational domain once the magnetic field is initialized (see below). We impose ceilings in the magnetization $\sigma=b^2/4\pi\rho$ and $\beta^{-1}=b^2/8\pi p$ (with $b$ the fluid frame magnetic field strength, $p$ the fluid frame pressure), injecting mass/energy when $\sigma>\sigma_{\rm max}=140$ and $\beta^{-1}>\beta^{-1}_{\rm max}=1200$. Additionally, if $\sigma>\sigma_{\rm max}$ for right or left states at cell interfaces, we use the Lax-Friedrichs numerical flux instead of the HLLE flux. We find $\sigma$ remains below $\sigma_{\rm max}$ throughout the computational domain. For models \texttt{B3P1.8} and \texttt{B3P1} (see Table ~\ref{tab:models}), we used Lax-Friedrichs numerical fluxes throughout the domain to increase numerical stability (as in \citealt{noble_direct_2009}).

As in Papers I and II, we use a one-moment (M0) approximation of the general-relativistic Boltzmann equation for neutrino transport adapted from \citet{radice_dynamical_2016}. Weak interactions include charged-current $\beta$-processes in emission and absorption, $e^\pm$ pair annihilation, plasmon decay, scattering on free nucleons and coherent scattering on heavy nuclei. Neutrino fluxes and average energies are evolved along null coordinate radial rays on a semi-spherical grid (polar angles $\theta<90^\circ$ corresponding to $z>0$ km, as we employ reflection symmetry) with 300 km radius centered on the proto-magnetar, with $n_r \times n_\theta \times n_\phi = 1300 \times 25 \times 50$ grid points (see Paper I for details). We compute the effective neutrino absorption at every hydro evolution time step. Though perhaps less accurate than a two-moment (M1) scheme, our M0 scheme is computationally much cheaper, making the current suite of simulations computationally feasible, and we do not expect the assumptions of our M0 scheme to appreciably affect our results. A comparison between a similar M0 and an M1 neutrino-transport scheme finds systematically higher $\nu_e$ and $\bar{\nu}_e$ luminosities in the M0 case by a factor $\approx\! 2$, whereas neutrino mean energies (except for higher species) and outflow properties (including nucleosynthesis) agree remarkably well; differences are typically within variations resulting from different assumptions within an M1 scheme such as the closure \citep{radice2022new}. Even directionally dependent properties in systems with strong departures from spherical symmetry agree remarkably well \citep{radice2022new}. On the other hand, a comparison between an M1 and a Monte-Carlo based neutrino-transport scheme in the neutron-star merger context \citep{foucart_monte-carlo_2020} finds that the M1 scheme systematically underpredicts $\bar{\nu}_e$ luminosities and leads to systematically more proton-rich outflows, which could make the M0 scheme more accurate than an M1 scheme in terms of $\nu_e$/$\bar{\nu}_e$ luminosities and outflow composition \citep{radice2022new}. Here, we focus solely on outflow dynamics (baryon-loading of field lines through $\nu_e$, $\bar{\nu}_e$) and acceleration of the wind by magnetic fields, which we do not expect to be significantly affected by the assumptions of our M0 scheme. Differences in $\nu_e$ and $\bar{\nu}_e$ luminosities and mean energies relative to more accurate transport schemes translate into slightly different times (proto-magnetar temperatures) during the early cooling phase.

The numerical grid is handled by the Carpet driver \citep{schnetter_evolutions_2004}, which implements Berger-Oliger mesh refinement with subcycling in time. Our Cartesian grid consists of a cubic base grid of radius $(32 \times 32 \times 16)\times 10^3$\,km containing 9 nested refinement levels, all centered on the proto-magnetar. For all proto-magnetar models with periods of $P>1$ ms, which do not exhibit significant distortion of their structure due to centrifugal forces, the innermost grid is a $31 \times 31 \times 31$\,km box with resolution $\Delta x\simeq 225$\,m. To account for oblateness and to better resolve equatorial outflows near the proto-magnetar surface, the innermost grid for rapidly rotating models (\texttt{P1, BP1, B.1P1, B.01P1}, $P \approx 1$\,ms) is a $(x,y,z) \simeq (60, 60, 30)$ km box, with the same $\Delta x\simeq225$\,m resolution. To save computational resources, we implement reflection symmetry across the equatorial ($z=0$ km) plane. As in Paper I and II, we have not found the use of reflection symmetry to have an impact on our results. A more detailed discussion on resolution, symmetries, and convergence is in Appendix~\ref{app:resolution}.

\begin{table*}[]
    \centering
\begin{tabular}{cccccccc}
 Model &
 $B_{\rm P}$ &
 $R_p/R_e$&
 $P$ &
 $\Omega/\Omega_{K}$ &
 $R_{\bar \nu_e}$ &
 $E_{\nu}$  [MeV] &
 $\langle L_{\nu} \rangle_{\rm iso}$ [$10^{51}$ erg s$^{-1}$]\\ 
  & 
 [$10^{15}$ G]&
 &
 [ms]&
 &
 [km] & 
 (pol , eq)$_{\bar \nu_e}$ ; (pol , eq)$_{\nu_e}$&
 (pol , eq)$_{\bar \nu_e}$ ; (pol , eq)$_{\nu_e}$ \\\hline \hline
\texttt{NRNM}    & 0   & 1.0   & n/a  & 0.00 & 10.3 & (20 , 20) ; (12.2 , 12.2) & (4.3 , 4.3) ;  (2.1 , 2.1)\\ 
\texttt{P18}     & 0   & 0.999 & 18   & 0.03 & 9.8  & (19.2 , 19.2) ; (12.5 , 12.7) & (3.4 , 3.7) ; (2.4 , 2.3) \\ 
\texttt{P1.8}    & 0   & 0.9   & 1.8  & 0.40 & 11.7 & (21.1 , 20.6) ; (11.5 , 11.5) & (2.8 , 2.4) ; (1.1 , 1.4) \\ 
\texttt{P1}      & 0   & 0.6   & 1.2  & 0.96 & 13.2 & (25.5 , 20.5) ; (13.3 , 11.5) & (2.2 , 2.6) ; (.5 , 5.6) \\ 
\hline 
\texttt{B3P18}   & 3   & 0.999 & 18.0 & 0.03 & 9.9  & (18.2 , 18.1) ; (13.1 , 13.2) & (2.1 , 2.5) ;  (2.0 , 2.5) \\ 
\texttt{B.6P1.8} & 0.6 & 0.9   & 1.8  & 0.40 & 10.9 & (19.4, 19.1) ; (12.1, 12.5) & (3.3 , 4.2) ; (2.1 , 2.7) \\ 
\texttt{B3P1.8}  & 3   & 0.9   & 1.8  & 0.40 & 11.8 & (18.5 , 18.5) ; (12.7 , 13.2) & (6.0 , 4.9) ; (3.9 , 3.0)\\ 
\texttt{B.6P1}   & 0.6 & 0.6   & 1.11 & 0.96 &  13.5 & (17.5, 18) ; (10.9 , 12.3) & (2.4 , 2.7) ; (1.5 , 1.9) \\ 
\texttt{B3P1}    & 3   & 0.6   & 1.11 & 0.96 & 16.2 & (18.9 , 19.4) ; (11.4 , 13.6) & (6.4 , 9.0) ; (4.5 , 3.7) \\ 
\end{tabular}
\caption{All proto-magnetar wind models assume a neutron star gravitational mass $1.4\,M_\odot$ and initial central temperature $20\,\mathrm{MeV}$. From left to right: initial poloidal surface magnetic field strength at the pole, initial ratio of polar to equatorial radius, initial spin period, equatorial surface angular velocity in units of the Keplerian value, equatorial radius of the neutrinosphere for electron anti-neutrinos, mean energies and isotropic equivalent luminosities of electron and anti-electron neutrinos at the pole (pol) and the equator (eq).}
\label{tab:models}
\end{table*}

All models are initialized with a rigidly rotating neutron star of gravitational mass 1.4 $M_\odot$ constructed with the \texttt{RNS} code \citep{stergioulas_comparing_1995} using the SFHo equation of state. As in Papers I and II, the radial profiles of the initial temperature and electron fraction are determined from beta-equilibrium as in \citet{Kaplan+14}. We assume an initial central temperature $T_c = 20$ MeV, similar to those of proto-neutron stars a few seconds after birth.  

A fixed metric is computed self-consistently by \texttt{RNS} together with the matter distribution at the initial time step; subsequent evolution proceeds with this fixed metric for computational efficiency. The spacetime does not deviate significantly from its initial data in time, as the proto-magnetar structure remains nearly constant and at most only a fraction of the proto-magnetar mass $<10^{-3} M_\odot$ is lost in matter outflows for the most extreme model over the course of the $\sim\!100$\,ms evolution.

As in Paper II, we evolve each model first without magnetic fields, at least until the neutrino-driven wind reaches an approximate steady-state within 100 km of the neutron star surface, as typically requires $t \sim 50\ms$. The angular profiles of neutrino luminosity and mean energy are self-consistently produced via diffusion through the non-spherical structure of the rotating proto-magnetar, and attain roughly steady state within $\lesssim 10$ ms of evolution. The polar and equatorial values of the isotropic luminosity and mean energies of the electron (anti-)neutrinos for each model are summarized in Table~\ref{tab:models}; the luminosities achieved $L_{\nu} \sim 10^{51}-10^{52}$ erg s$^{-1}$ are typical of those of PNSs formed in core-collapse SNe \citep{Pons+99} or neutron star merger remnants \citep{Dessart+09} in the first few seconds after forming.  


\begin{figure}
    \centering
    \includegraphics[width=1.0\linewidth]{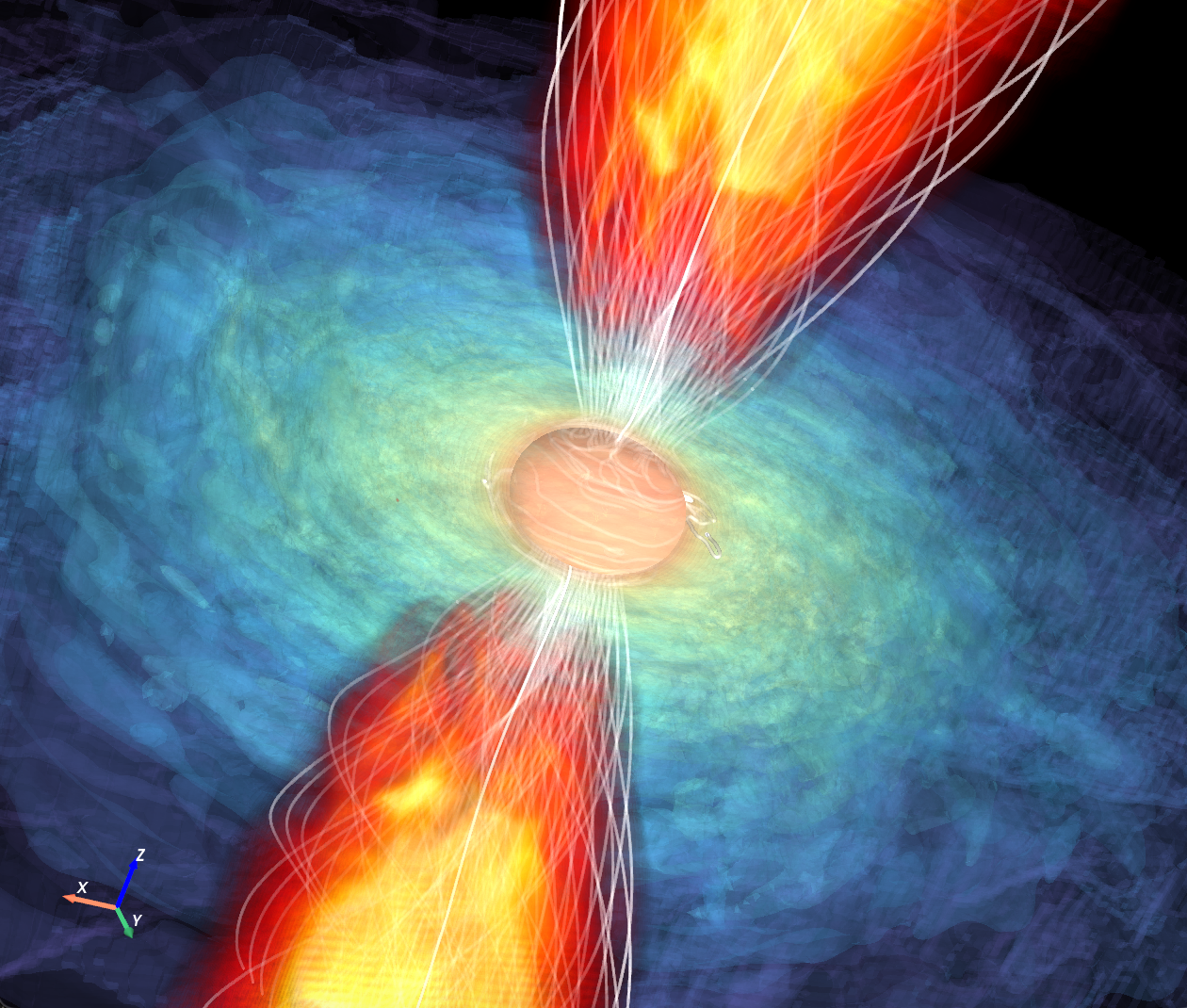}
    \caption{Volume rendering of a rapidly rotating proto-magnetar launching powerful equatorial winds and a magnetized jet as simulated in our \texttt{B3P1} model. Density isosurfaces are represented in light red ($\rho\approx 10^{14}\,{\rm g}{\,\rm cm}^{-3}$) and blue ($\rho\approx 10^{9}\,{\rm g}{\,\rm cm}^{-3}$), whereas magnetization is represented in red ($\sigma\approx 1$) and yellow ($\sigma\approx 100$) contours. White lines show polar magnetic field lines.}
    \label{fig:3D}
\end{figure}

After a steady state is reached for the purely $\nu-$hydrodynamical evolution, we initialize our simulations with a purely poloidal dipole field, where $B_{\rm P}$ is the initial surface poloidal magnetic field strength at the pole ($\theta = 0^\circ$). Each simulation is then subsequently evolved towards an approximate steady-state as differential rotation develops near the stellar surface, where the wind is loaded with mass.  Purely poloidal magnetic fields inside a star are known to be unstable in ideal MHD (e.g., \citealt{ciolfi_instability-driven_2011, sur_long-term_grmhd_2022}), typically leading to reconfiguration of the field structure over the Alfv\'en crossing time of the star (though rapid rotation can slow down the growth of the instability; e.g., \citealt{braithwaite_stability_2007, venturi_rotation_2025}). 
Although we do see moderate evidence for interior field evolution in our simulations, these changes occur slowly during the limited time window that we evaluate the steady-state wind properties, over which the surface poloidal field strength is approximately constant.  As remarked earlier, although the strong fields we assume are likely generated by a dynamo within the star, such field amplification is likely to take place over a much shorter timescale (e.g., the MRI grows on the rotational period of milliseconds) than the neutrino cooling timescale of several seconds we are trying to model.  
For these reasons, we focus exclusively on steady-state wind properties rather than transient magnetic evolution.

The magnetosphere outside the star experiences occasional brief flux eruptions, driven by neutrino heating in the closed zone, which opens field lines and expels matter (\citealt{thompson_magnetic_2003,Prasanna+22}; Paper II). Despite these transient events, the magnetic field structure within the wind (the surface magnetic field strength at the proto-magnetar surface $B_{\rm P}(R_\nu)$ in particular, which determines the magnetic flux of the wind) reaches steady-state evolution within approximately 50 ms and we diagnose wind properties from this point onward.

The parameters and initial conditions for our suite of simulations are summarized in Table~\ref{tab:models}.  Each model is labeled according to its magnetic field strength at the polar cap $B_{\rm P} \approx 10^{13}-3\E{15}$\,G and intial spin period $P\approx 1-18\ms$, the latter corresponding to polar-to-equatorial radii from $R_p/R_e = 0.6$ to $0.999$.  For comparison, we also evolve a non-rotating, non-magnetized wind model, labeled \texttt{NRNM}.

\begin{figure*}
    \centering
    \includegraphics[trim={0cm 0cm 0cm 0cm},clip,width=1.1\linewidth]{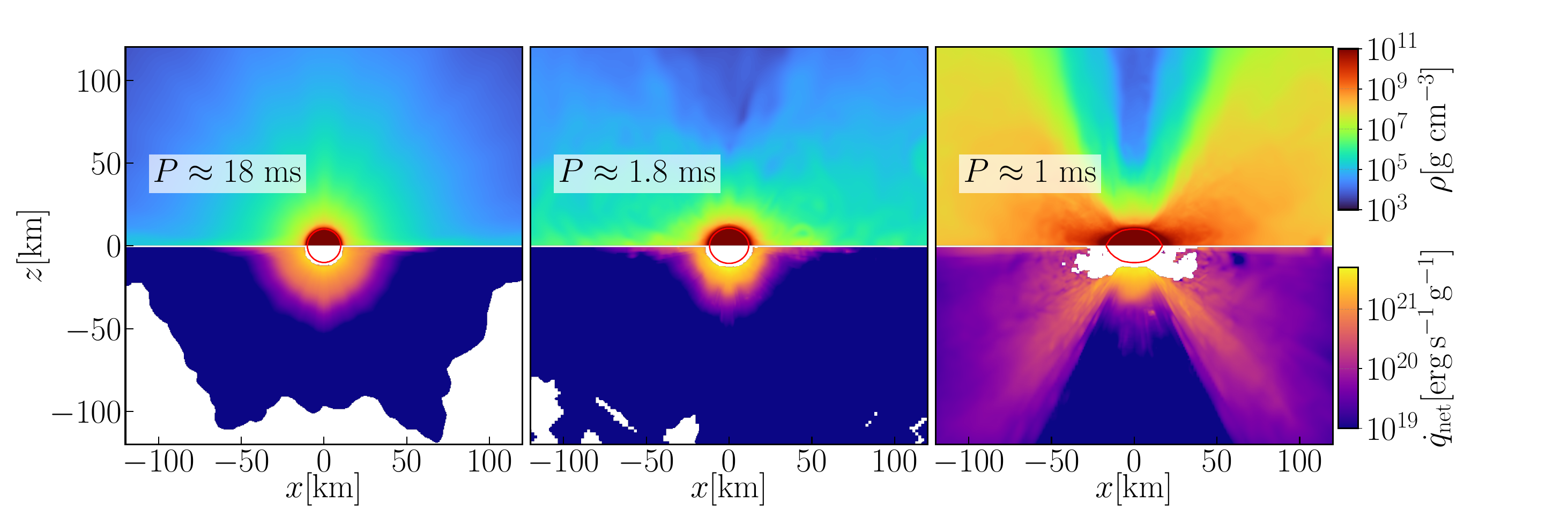}
    \includegraphics[trim={0cm 0cm 0cm 1.2cm},clip,width=1.1\linewidth]{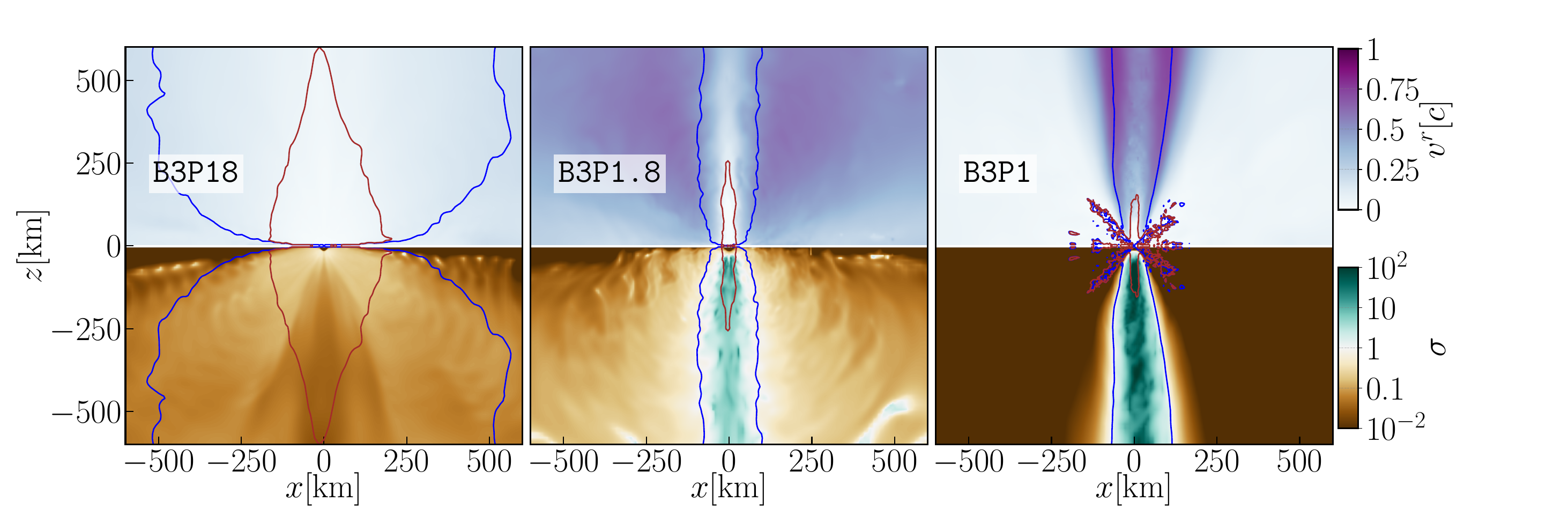}
    \vspace{-.75cm}
    \caption{We show 2D cross-sections of the $y=0$ plane time-averaged over $\approx 5$ ms for our most strongly magnetized models with $B_{\rm P} =3\E{15}$ G, for three different spin periods (left to right) $P=18$ ms, $P=1.8$ ms, and $P=1$ ms. \textit{Top row}: Rest-mass density $\rho$ ($z>0$) and net neutrino heating rate $\dot q_{\rm net} \equiv \dot q_+ - \dot q_-$ ($z<0$).  White shading represents regions of net cooling, and red contours denote the electron anti-neutrino sphere ($\tau_{
        \bar \nu_e}\simeq 2/3$). \textit{Bottom row}: Radial 3-velocity $v^r$ ($z>0$) and magnetization $\sigma = b^2/4\pi\rho$ ($z<0$).  The blue (outer) contour denotes the Alfv\'{e}n surface (where the radial Alfv\'{e}n velocity $v_A^r=v^r$), whereas the brown (inner) contour denotes the slow magnetosonic surface.}
        \label{fig:cross-sections}
\end{figure*}

\section{Results}
\label{sec:results}

\begin{figure}
    \centering
        \text{\large $r=500$ km}
        \hspace*{-.1cm}\includegraphics[trim={0cm .35cm 0cm 0cm},clip,width=.95\linewidth]{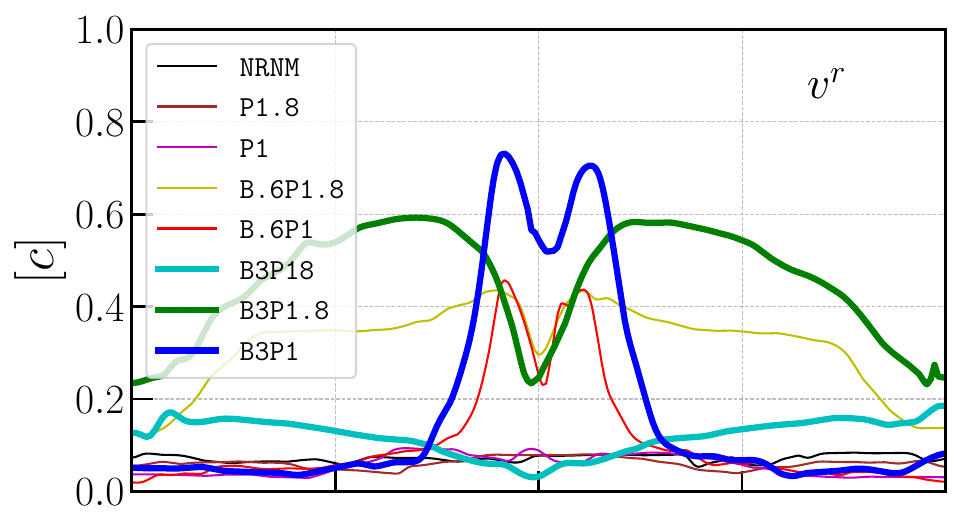}        \hspace*{-.08cm}\includegraphics[trim={0cm .25cm -.5cm .2cm},clip,width=1.01\linewidth]{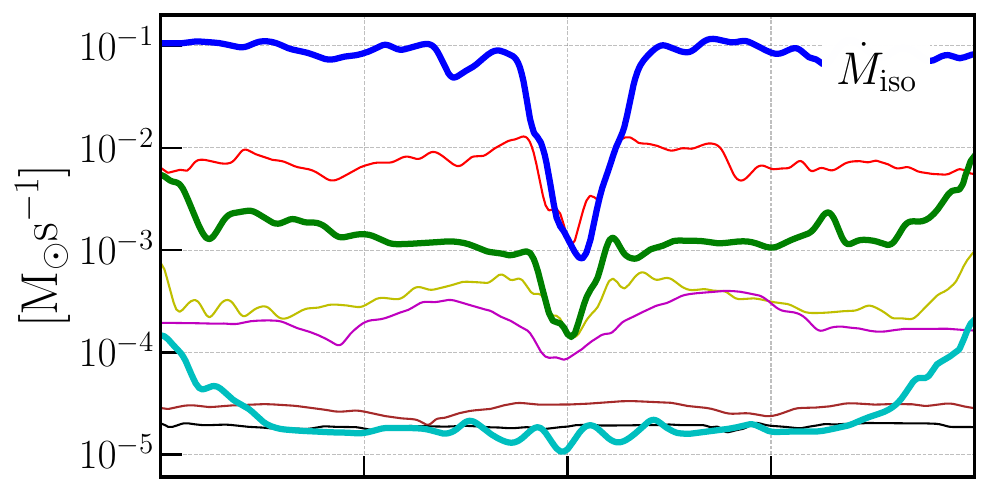}        \includegraphics[trim={-.25cm .35cm .2cm 0cm},clip,width=.92\linewidth]{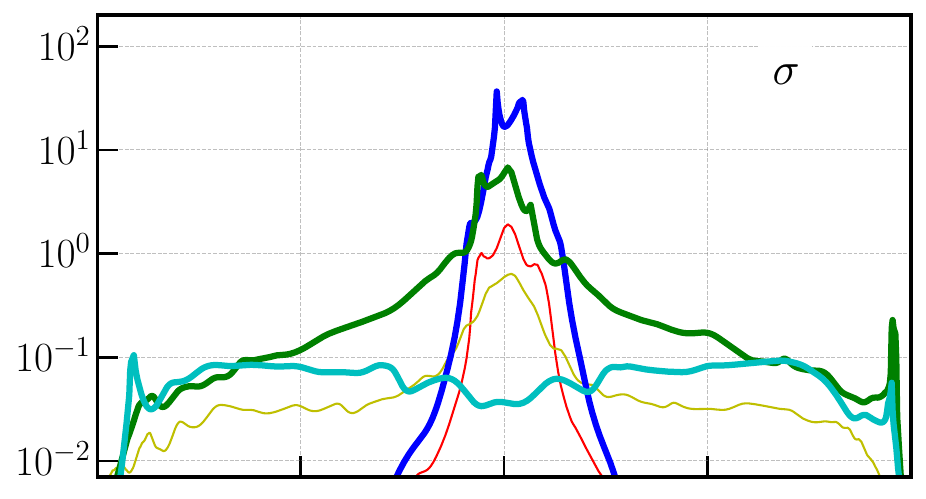}
        \includegraphics[trim={0cm .24cm 0cm .1cm},clip,width=1.00\linewidth]{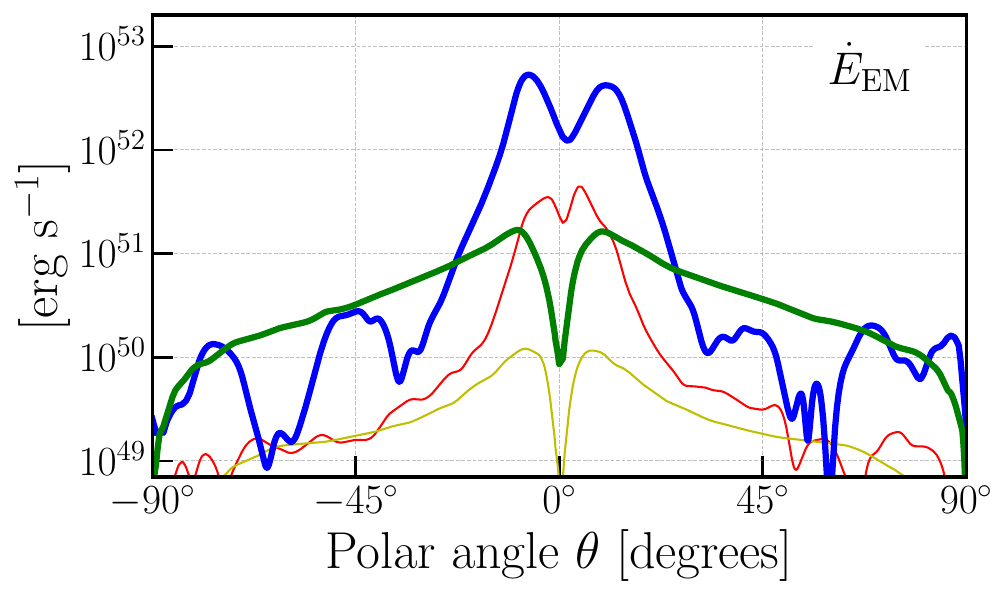}
    \caption{Profiles of (from top to bottom) radial 3-velocity $v^r$, isotropic-equivalent mass outflow rate $\dot M_{\rm iso}$, magnetization $\sigma$, and isotropic-equivalent electromagnetic power $\dot E_{\rm EM}$ in the magnetar outflow as a function of polar angle $\theta$, as determined through a sphere of radius $r=500$ km from 2D slices time-averaged over an interval $\sim5$ ms after the wind has achieved a quasi steady-state.}
    \label{fig:angular}
\end{figure}

\begin{figure*}

        \centering
    \includegraphics[trim={1.8cm .55cm 1.5cm 1.4cm},clip,width=.47\linewidth]{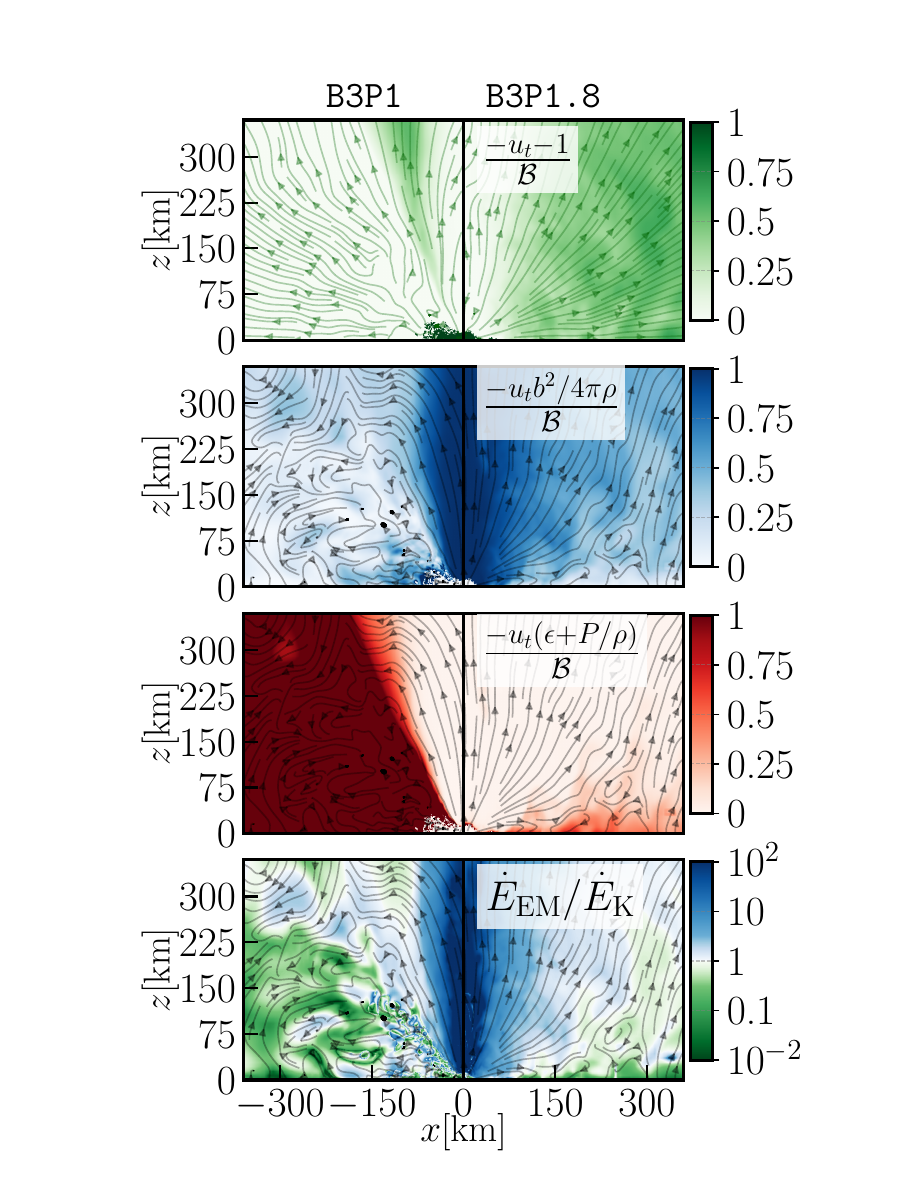}
    \includegraphics[trim={.6cm 1.5cm 2cm 0cm},clip,width=.52\linewidth]{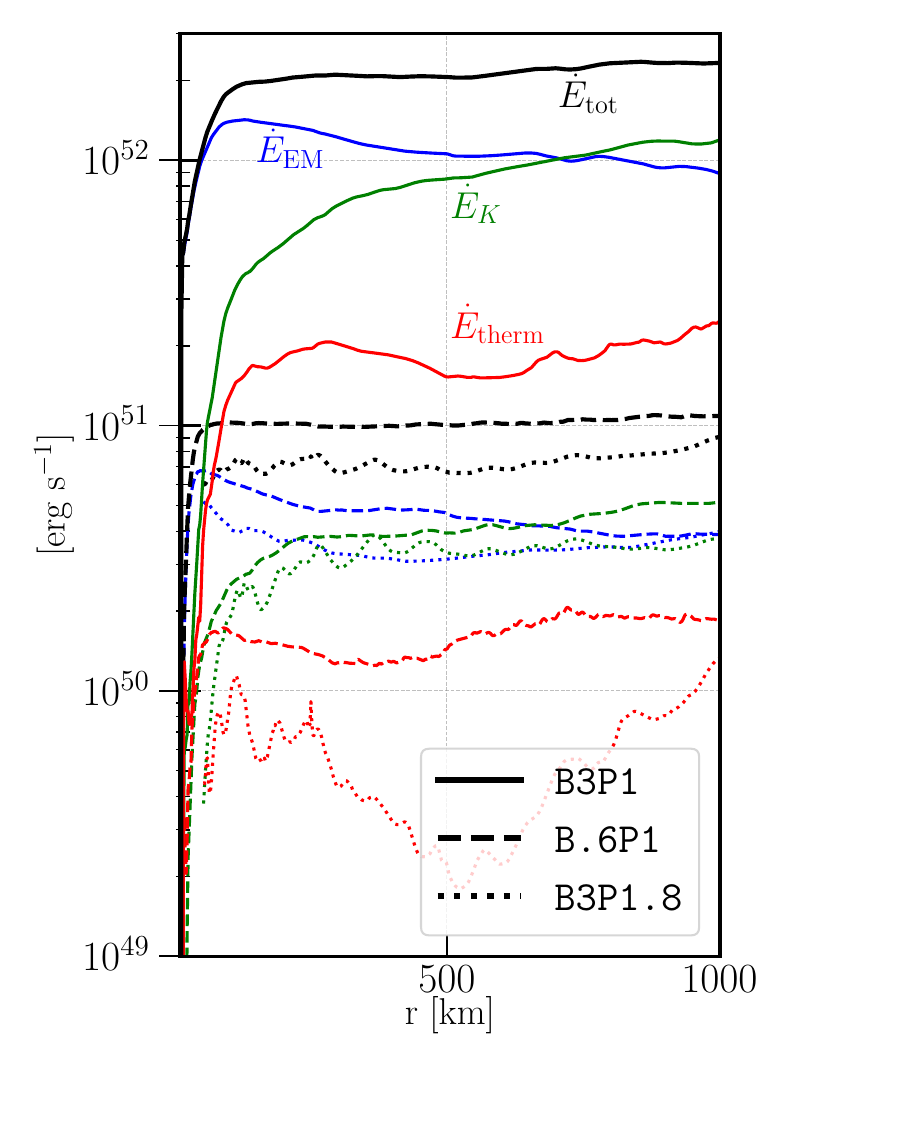}

    \caption{\textit{Left:} Meridional cross sections showing steady-state properties (averaged over $\approx 5$ ms) of the wind models \texttt{B3P1} (left) and \texttt{B3P1.8} (right). The top three rows (from top to bottom) show the fractional contributions of the specific kinetic energy ($\approx -u_t-1$ at large distances), magnetic energy ($-u_tb^2/4\pi\rho$), and thermal enthalpy ($-u_t(\epsilon +P/\rho)$), to the total specific wind energy $\mathcal{B}$ (defined in Eq.~\eqref{eq:bern}). The bottom row shows the ratio of electromagnetic $\dot E_{\rm EM}$ to kinetic $\dot E_{\rm K} \equiv (W-1) \dot M_{\rm iso}$ power. The top row shows fluid velocity field lines, whereas the bottom three rows show the magnetic field lines.
    \textit{Right:} Angle-averaged radial profiles of isotropic-equivalent luminosities along the $y=0$ plane for three models: \texttt{B3P1} (solid, $0^\circ \le \theta \le 30^\circ$), \texttt{B.6P1} (dashed, $0^\circ \le \theta \le 30^\circ$), \texttt{B3P1.8} (dotted, $0^\circ \le \theta \le 90^\circ$). We show separately the Poynting luminosity (blue), kinetic luminosity (green), thermal luminosity (red), and their sum (black). 
    }
    \label{fig:Econv}
\end{figure*}

Figure~\ref{fig:3D} shows a volume rendering of the jet and wind structure of our most highly spinning and most strongly magnetized model \texttt{B3P1}, whereas Fig.~\ref{fig:cross-sections} explores in more detail 2D cross-sections through the rotational axis illustrating the steady-state structure of the wind's radial velocity $v^r$, magnetization $\sigma$, and net specific neutrino-heating rate $\dot q_{\rm net}$ (see Eq.~(9) of Paper I for details), for all most strongly magnetized models ($B_{\rm P} \approx 3\times 10^{15}$ G) for three different spin periods ($P \approx 18, 1.8, 1$ ms). For the wider suite of models, Fig.~\ref{fig:angular} shows similarly time-averaged steady-state profiles of the polar angular dependence of $v^r$, isotropic-equivalent mass-loss rate $\dot M_{\rm iso}$, magnetization $\sigma \equiv b^2/4\pi\rho$, and isotropic-equivalent electromagnetic power $\dot E_{\rm EM}$. Fig.~\ref{fig:Econv} shows 2D cross-sections (left panel) and radial profiles (right panel) of kinetic ($\dot E_{\rm K}$), electromagnetic ($\dot E_{\rm EM}$) and thermal ($\dot E_{\rm therm}$) components of outflow luminosities for several models. The left panel of Fig.~\ref{fig:jet_param} shows the cumulative power of these components contained in outflows above a given magnetization for model \texttt{B3P1}. The right panel of Fig.~\ref{fig:jet_param} shows the total isotropic-equivalent energy versus isotropic-equivalent kinetic energy for our models, in comparison to observations of GRB with associated kilonovae. 
For a non-magnetized, slowly-rotating neutron star (model \texttt{NRNM}), the outflow is approximately spherically symmetric with a small mass-loss rate $\dot{M}_{\rm iso} \approx 2\times 10^{-5}M_{\odot}$ s$^{-1}$ and small velocity $\lesssim 0.1$ c (Fig.~\ref{fig:angular}), both consistent with standard PNS winds driven by thermal pressure from neutrino heating (e.g., \citealt{Qian&Woosley96,Thompson+01}).  For stronger magnetic fields and faster spin rates, $\dot{M}_{\rm iso}$ rises relative to the \texttt{NRNM} baseline, particularly in the equatorial regions ($\theta \sim \pm 90^{\circ}$) where centrifugal effects become important; these become sizable for spin periods $P = 2\pi/\Omega \lesssim P_{\rm cf} \approx 2$ ms (Eq.~\eqref{eq:Pcf}), for which the rotational velocity near the equator, $v^{\phi} \approx R_{\nu} \Omega \approx 0.1(P_{\rm ms}/2{\rm \, ms})^{-1}\,c$, exceeds the surface sound speed $c_{\rm s} \sim 0.1 c$, the latter set by the balance of neutrino heating and cooling (Eq.~\eqref{eq:Pcf}).  

The combination of magnetic fields and rotation greatly boost the wind power, with both the magnetization and asymptotic velocity rising with increasing $B_{\rm P}$ and decreasing $P$.  Though magneto-rotational effects come to dominate the wind acceleration (e.g., \citealt{Belcher&MacGregor76}), for most models the outflow remains non-relativistic ($u^{r}, \sigma < 1$) along all directions because of the baryon loading from neutrino heating near the surface. Such moderately magnetized and/or modestly rapidly spinning magnetars contribute to the energy of the surrounding supernova/kilonova ejecta (e.g., \citealt{thompson_magnetar_2004}), but they are not capable of generating relativistic GRB jets, at least within the first few seconds necessary to power a short GRB.

For our most rapidly spinning and highly magnetized models, a qualitatively different picture emerges.  As shown in Fig.~\ref{fig:angular}, the outflow can be divided into three distinct angular regions: a wide equatorial belt ($\theta \gtrsim 30^\circ$), a narrow jet-like polar region along the rotation axis ($\theta \lesssim 10^\circ$), and a transition region at intermediate latitudes ($\theta \approx 10-30^\circ$).  

In the equatorial zone, $\dot{M}_{\rm iso}$ peaks at a value several orders of magnitude higher than the \texttt{NRNM} model (Fig.~\ref{fig:angular}, second panel from the top).  This centrifugal enhancement results from an increase in the density scale-height near the neutron star surface, to which the mass-flux is proportional (e.g., \citealt{Qian&Woosley96}), due to the reduction in the effective gravitational acceleration caused by co-rotation of the atmosphere enforced by the strong magnetic field (e.g., \citealt{Lamers&Cassinelli99,metzger_proto-neutron_2007}).  This effect is illustrated by the expanded radial extent of the region of net neutrino heating near the equator in model \texttt{B3P1} (Fig.~\ref{fig:cross-sections}, bottom half of the top row), as compared to the slower rotating models.    

The wind is accelerated to its highest velocities along the intermediate range of angles (Fig.~\ref{fig:angular}, top panel; top half of the bottom row in Fig.~\ref{fig:cross-sections}). Figure~\ref{fig:Econv} shows how the radial profile of the wind energy is partitioned into separate magnetic, kinetic and thermal (enthalpy) contributions. For model \texttt{B3P1.8}, the outflowing matter is initially magnetically dominated within a cylindrical radius of $\simeq 50$ km, before becoming kinetically dominated on larger scales. The fact that this conversion from magnetic to kinetic energy occurs with minimal thermal energy indicates that centrifugal slinging, instead of magnetic dissipation (e.g., \citealt{Drenkhahn&Spruit02}), is responsible for the wind's acceleration (e.g., \citealt{spruit_theory_2008}; their Sec.~9.3.2).  Our fastest spinning model \texttt{B3P1} exhibits similar energy gradients within a narrow range of intermediate angles $10^\circ \lesssim \theta \lesssim 30^\circ$ (left half of the left panels in Fig.~\ref{fig:Econv}), where the wind achieves its highest velocity (Fig.~\ref{fig:angular}, top panel). The efficiency of magnetic-to-kinetic energy conversion in the polar outflows exceeds expectations for the asymptotic energy partitioning $\dot{E}_{\rm EM} \simeq 2 \dot{E}_{\rm KE}$ in non-relativistic, unconfined (split-monopole) outflows (e.g., \citealt{Lamers&Cassinelli99}), as illustrated in the bottom row of the left panel of Fig.~\ref{fig:Econv}. This may indicate that external confinement and polar collimation by the dense equatorial regions of the wind play a role in accelerating the polar outflows, as discussed further below. For models with more modest spin periods and less elevated $\dot{M}_{\rm iso}$ in the equatorial regions such as \texttt{B3P1.8}, the sub-relativistic winds indeed attain the expected asymptotic partitioning $\dot{E}_{\rm EM} \simeq 2 \dot{E}_{\rm KE}$ over a broad range of intermediate angles, whereas for \texttt{B3P1} this region is `compressed' to a narrow angular range around the polar jet-like region (Fig.~\ref{fig:Econv}, bottom left panel).

For the most rapidly spinning model \texttt{B3P1}, the polar most field lines define a distinct jet region that reaches a peak magnetization $\sigma \sim 30-100$ (Fig.~\ref{fig:cross-sections}), much higher than at lower latitudes in the wind.  As we describe in Appendix \ref{app:analytic}, the polar field lines threading the jet are much cleaner than those at lower latitudes, because the polar outflows do not experience centrifugal enhancement of their mass-loss rate as a result of the smaller rotational lever arm of their footpoints within a modest angle $\theta_{\rm j} \lesssim 20^{\circ}$ close to the pole. {Moreover, rapidly spinning models ($P \approx 1, 1.8$ ms) show $\sim2$ orders of magnitude lower rest-mass density in the polar funnel than the slower $P \approx 18$ ms model (Fig.~\ref{fig:cross-sections}, top panel), because fluid is driven away from the rotational axis by strong centrifugal forces. Since the magnetic flux in the polar funnel is roughly constant across the strongly magnetized models, the magnetization $\sigma$ is significantly boosted for the rapidly rotating models (Fig.~\ref{fig:cross-sections}, bottom panels). On the other hand, the mass loss rate increases with rotation rate even in the polar region 
(Fig.~\ref{fig:angular}, second panel from top) because collimation dramatically enhances the radial wind velocity (Fig.~\ref{fig:angular}, top panel), offsetting the lower density.}

Whereas self-collimation due to hoop stress is possible in the case of non-relativistic magnetized outflows (e.g., \citealt{spruit_theory_2008}), high $\sigma \gg 1$ ultra-relativistic outflows are known to be incapable of self-collimation (e.g., \citealt{Bogovalov01,narayan_stability_2009}), thus relying on collimation by an external medium such as the supernova or kilonova ejecta (e.g., \citealt{Komissarov+09,bromberg_propagation_2011}).  The massive equatorial outflows help to collimate the more highly magnetized polar outflows across a narrower jet half-opening angle $\theta_{j,\rm col}\simeq 8^\circ$ (which we define to be roughly the polar angle within which there are $\sigma \gtrsim 10$ outflows) than were the field lines of the outflow to expand purely radially off the surface (see the angular profiles of $\sigma$ in the third panel from the top in Fig.~\ref{fig:angular}). The ``hollowed out'' angular structure of the four-velocity $u^{r}(\theta)$ seen in our simulations qualitatively resembles those seen in GRMHD simulations of black hole jets (e.g., \citealt{Tchekhovskoy+10}).



In Appendix \ref{app:analytic}, we derive analytic estimates for the magnetization of the proto-magnetar outflows.  We separately consider: (1) the magnetization of the bulk wind, $\sigma_{\rm wind}$ (Eq.~\eqref{eq:sigma_wind}), which is dominated by slow but powerful outflows from low latitudes near the last closed field lines that experience the largest centrifugal effects on their mass-loading; (2) the higher magnetization of the polar jet, $\sigma_{\rm j}$ (Eq.~\eqref{eq:sigma_jet}), which arise from high latitudes near the neutron star polar cap that experience a much lower isotropic baryon loading rate close to the \texttt{NRNM} value.  Figure~\ref{fig:sigma} in Appendix \ref{app:analytic} shows contours of $\sigma_{\rm wind}$, $\sigma_{\rm j}$ in the space of $B_{\rm P}$ and $P$, calculated for fiducial assumptions and a fixed neutrino luminosity $L_{\bar \nu_e}=6\E{51}$ erg s$^{-1}$.  Our fastest spinning models with $B_{\rm P} =3\E{15}$ G and $P = 1(1.8)$\,ms lie in a region with $\sigma_{\rm j} \gtrsim 50(20)$, broadly consistent with the peak values of $\sigma$ attained in the jet core for these models (Fig.~\ref{fig:angular}, third panel). 

A few in $10^{52}$ erg s$^{-1}$ of total isotropic-equivalent power is contained in the jet for model \texttt{B3P1} (black solid line in the right panel of Fig.~\ref{fig:Econv}).
This is significantly larger than the $\simeq 5\E{50}$ erg s$^{-1}$ predicted by the standard dipole formula (relativistic outflow) 
for a neutron star of similar magnetic field strength and rotation rate. This is because the winds simulated here are closer to a split-monopole configuration and the non-relativistic limit \citep{thompson_magnetar_2004}, as well as shaped by multidimensional effects.

\section{Discussion: Implications for GRBs from Young Millisecond Magnetars}
\label{sec:discussion}

To power a short GRB jet requires that the central engine place significant energy into a collimated outflow with a bulk Lorentz factor $\Gamma > 10-100$ (e.g., \citealt{Lithwick&Sari01,nakar_short-hard_2007}).  As described in the previous section, our most strongly magnetized, rapidly spinning proto-magnetar models indeed achieve $\sigma \simeq \Gamma_{\infty} \sim 30-100$,\footnote{Several works address how such a Poynting-dominated magnetized jet on small scales close to the central engine might convert its magnetic energy into bulk kinetic energy to achieve $\Gamma \simeq \Gamma_{\infty}$ on the much larger scales relevant to GRB prompt emission (e.g., \citealt{Drenkhahn&Spruit02,Komissarov+09,Granot+10}), and in any case similar acceleration requirements apply to black hole jets powered by the \citet{Blandford&Znajek77} mechanism.} thus satisfying this key requirement (see Appendix~\ref{app:bern} for a discussion of $\Gamma_{\infty}$).

\begin{figure*}
    \centering
     \includegraphics[trim={0cm -.2cm 0cm 0cm},clip,width=.52\linewidth]{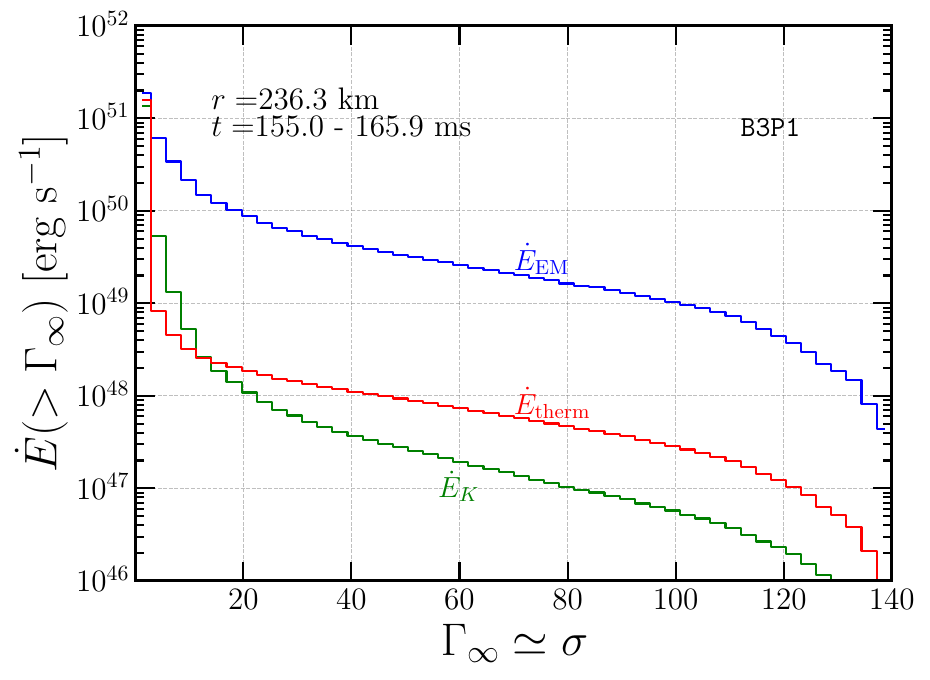}        \includegraphics[width=.465\linewidth]{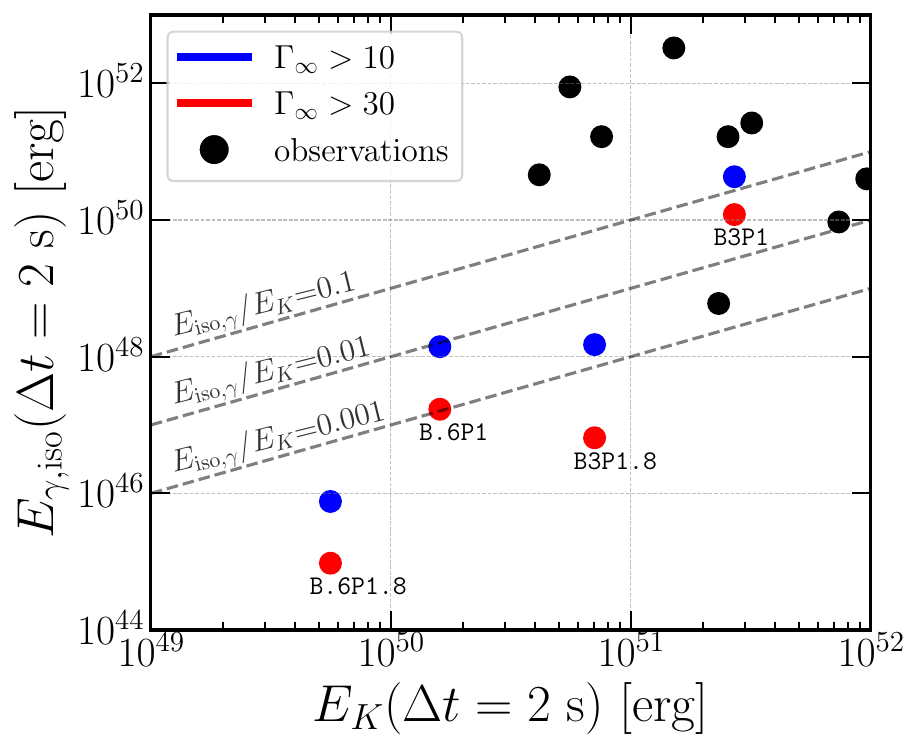}
    \vspace{-.3cm}
    \caption{\textit{Left}: Cumulative distribution of wind luminosity $\dot E$ in material above a given value of $\Gamma_\infty \simeq \sigma$ (Eq.~\eqref{eq:gamma_inf}), for the most rapidly spinning, highly magnetized model (\texttt{B3P1}).  The Poynting ($\dot E_{\rm EM}$, blue), thermal ($\dot E_{\rm therm}$ red), and kinetic ($\dot E_{\rm K}$, green) luminosity of the wind are measured at a spherical shell of radius $r\simeq 236$ km, between $t\simeq 155-166$ ms after magnetic field initialization. {\it Right:} Total isotropic-equivalent energy carried by the outflow over a duration $\Delta t=2\,\mathrm{s}$, in material with asymptotic Lorentz factor $\Gamma_\infty > 10$ ($\Gamma_\infty > 30$) shown with blue (red) points, vs. total isotropic-equivalent kinetic energy of the outflow $E_K (\Delta t=2~\rm s)$, as measured from cumulative distributions shown in the left panel. Most of the energy in fluid with $\Gamma_\infty\lesssim10$ resides in sub or moderately-relativistic wind components that would likely couple to the surrounding supernova or kilonova ejecta. Black points show inferred isotropic-equivalent energies of gamma-ray emission $E_{\gamma,\rm iso}$ and associated kilonovae $E_K$ from short GRBs compiled by \cite{rastinejad_kn_grb_2025}, rescaled to a common duration of $2\,\mathrm{s}$ and assuming a jet half-opening angle $\theta_{j,\rm col} \simeq 8^\circ$ to match those found from the strongly magnetized, rapidly spinning wind models.}
        \label{fig:jet_param}
\end{figure*}

The left panel of Fig.~\ref{fig:jet_param} shows, for our most rapidly spinning model \texttt{B3P1}, the cumulative outflow power above a given $\Gamma_\infty \sim \sigma$.  Because $\sigma(\theta)$ monotonically decreases with polar angle $\theta$, this distribution can also be mapped into an angular power pattern (Fig.~\ref{fig:angular}, bottom panel).  The proto-magnetar jet releases $\dot{E} \approx \dot{E}_{\rm EM} \sim 10^{49}-10^{50}$ erg s$^{-1}$ in material with $\Gamma_\infty \sim 10-100$, in principle sufficiently luminous and clean to power a short-duration gamma-ray burst, for which typical beaming-corrected gamma-ray luminosities reside in this same range, $L_{\gamma} \sim 10^{49}-10^{51}$ erg s$^{-1}$ (e.g., \citealt{nakar_short-hard_2007}). The jet half-opening angle of $\simeq 8^\circ$ is in good agreement with estimates of the typical half-opening angles of observed short and long GRBs of $(16\pm 10) ^\circ$ \citep{fong_decade_2015} and ${6^\circ}_{-2}^{+4}$ \citep{goldstein_estimating_2016}, respectively.

Even for models in which the polar jet achieves $\sigma \gg 1$, more heavily mass-loaded outflows from the jet sheath and equatorial wind at greater angles off the rotational axis achieve only sub-relativistic speeds, $ v \sim \sigma^{1/3} c \sim 0.05-0.5 c$ with magnetic and kinetic luminosities in rough equipartition (Fig.~\ref{fig:Econv} right panel and bottom left panel). These sub-relativistic outflows carry the bulk of the wind's energy and hence still may play a crucial role in shaping the electromagnetic signatures by boosting the kinetic energy of the supernova (e.g., \citealt{thompson_magnetar_2004}) or kilonova (e.g., \citealt{metzger_magnetar_2018, combi_jet_2023}) ejecta.

The neutrino luminosity of a proto-neutron star decreases on the Kelvin-Helmholtz cooling timescale of $\sim\!1$\,s (e.g., \citealt{Roberts&Reddy17}), and therefore the wind properties achieved by model \texttt{B3P1} will change over similar timescales. Nevertheless, unless and until it collapses into a black hole\footnote{A collapse timescale of $\sim$seconds might occur if the remnant destabilizes due to the effects of neutrino cooling on thermal support of the PNS \citep{Kaplan+14} or by mass accretion through a disk (not modeled here) which drives the object above the maximum stable neutron star mass \citep{margalit_angular_2022}. 
}, the magnetar will continue to produce a jet, albeit one that is likely to become more relativistic and less collimated as neutrino-driven mass-loss and centrifugal effects abate.  Many short GRBs exhibit temporally extended X-rays lasting tens or hundreds of seconds (e.g., \citealt{Perley+09}). Although usually less luminous than the initial ``short'' prompt spike, this distinct emission component could also arise from the proto-magnetar jet if its opening angle widens in time (e.g., \citealt{Metzger+08a}), decreasing the isotropic equivalent luminosity. Collimation of the magnetar jet is both a function of the intrinsic wind structure (as studied here) as well as by the external medium on larger scales (e.g., \citealt{bromberg_propagation_2011}). Jet collimation may therefore persist longer for magnetars formed in core-collapse supernovae embedded within the confining progenitor envelope, whereas merger-produced magnetars are surrounded by only the merger ejecta; within the magnetar picture, mergers thus appear to be a more favored origin for extended emission short GRBs.

Relating to the above argument, the right panel of Fig.~\ref{fig:jet_param} shows the wind energy released by a hypothetical magnetar with outflow luminosities predicted by each of our models, assuming a wind/jet duration of 2 seconds, as motivated by the short GRB case.  We divide the energy of each magnetar outflow model into its ultra-relativistic jet ($\Gamma_{\infty} = \sigma >10 ~(30)$ with blue (red) points, respectively) as well as the sub-relativistic wind kinetic energy, which would instead likely be imparted to the kilonova ejecta (e.g., \citealt{yu_bright_2013, ciolfi_short_2015, metzger_magnetar_2018, combi_jet_2023}).  For comparison, we show with black points the inferred kilonova and (beaming-corrected) gamma-ray energies from a sample of short GRBs with measured kilonova counterparts, as compiled by \citet{rastinejad_kn_grb_2025}, rescaled to the same duration of 2 seconds.  Our most rapidly-spinning model \texttt{B3P1} is seen to overlap with the GRB/kilonova sample.  Because spin periods of around 1 ms are indeed similar to those expected of post-merger (and, potentially, AIC) remnants (e.g., \citealt{Radice+18}), we conclude that a short-lived proto-magnetar phase could contribute to their full range of electromagnetic counterparts.  

There are several caveats worth noting to the simplified analysis presented here. Neutron star merger and AIC events intrinsically possess, or develop through redistribution of angular momentum within the remnant, a significantly massive disk surrounding the remnant, which we do not model in our simulations.  The outflows from such a disk will also contribute (e.g., \citealt{siegel_three-dimensional_2018}), if not dominate, over that from magnetar winds, in which case the magnetar kinetic energy contribution in the right panel of Fig.~\ref{fig:jet_param} is a lower limit.  The hyper- or supramassive remnants of neutron star mergers also possess a higher mass $\gtrsim 2 M_{\odot}$ than assumed in our models; this will act to reduce the wind mass loss rate $\dot{M} \propto M^{-2}$ (Eq.~\eqref{eq:Mdotsph}), and hence boost $\sigma \propto 1/\dot{M} \propto M^{2}$, by a factor of several compared to our simulations which assume $M = 1.4M_{\odot}$.

Another key question is the long term evolution of the magnetar outflow and its implications for high-energy emission.  In post-merger or core-collapse environments, the magnetar may accrete enough matter from the surrounding star or disk to collapse to a black hole (e.g., \citealt{margalit_angular_2022}); however, this may not always be the case (e.g., \citealt{dessart_proto-neutron_2008}), and in fact is very unlikely in white dwarf AIC scenarios, leaving a long-lived or indefinitely stable magnetar.  As the magnetar spins down and cools, neutrino-driven mass-loss abates, enabling a greater and greater fraction of the field lines leaving the polar cap to achieve high $\sigma,$ capable of powering a relativistic outflow (e.g., \citealt{Metzger+11a}).  At the same time, spin-down will render the slower centrifugally-boosted outflow less powerful, providing weaker collimation of the high-$\sigma$ wind/jet.

Thus, over the course of seconds to minutes, the magnetar jet may become more powerful but wider in angle, $\theta_{\rm jet}$. Viewers along the spin axis may see high-energy emission from both phases, with the detailed light curve shape depending on the evolution of the isotropic luminosity,
$L_{\rm iso} \propto \dot{E}(\sigma > 100)/\theta_{\rm jet}^{2}$, but conceivably comprised of distinct ``prompt'' ($\lesssim$ seconds) and ``extended'' ($\sim$ minutes) emission phases. We emphasize that a widening jet does not by itself guarantee an extended-emission plateau: if $\theta_{\rm jet}$ grows too rapidly compared to the evolution of $\dot{E}$, the observed isotropic-equivalent luminosity would instead decline continuously. Nevertheless, because observed extended emission is typically longer-lived and less luminous than the initial spike, with a comparable radiated energy budget (e.g., \citealt{Perley+09}), gradual changes in both jet power and collimation remain a plausible ingredient of such models.

A separate issue for any long-lived magnetar model is the fate of the remaining rotational energy, since comparable energy injection has not yet been clearly observed in kilonova ejecta or in late-time jet afterglows (e.g., \citealt{ghosh_merger_emission_2024}). One possible way to reduce the available rotational reservoir is through torques exerted by a surrounding disk: the disk can in principle ``brake'' the magnetar without depositing all of the extracted spin energy into the ejecta or jet, provided that the disk radiates much of the received energy efficiently in neutrinos. As discussed in \citet{metzger_fallback_2018}, disk torques and magnetar spin-down can balance at an equilibrium spin period that may be as low as $\lesssim 3$ ms as the accretion rate drops, at times later than simulated here. This would reduce the ultimate rotational energy released by the magnetar to only a few $10^{51}$ erg. However, the fate of the remaining rotational energy remains an important uncertainty for any model in which the magnetar survives for $\sim$tens of seconds or longer.

\section{Conclusions}
\label{sec:conclusions}
We have presented the first three-dimensional general-relativistic magnetohydrodynamic simulations of neutrino-heated winds from rapidly rotating, strongly magnetized proto-magnetars, with the goal of assessing whether such objects can self-consistently launch relativistic jets within the first few seconds after formation. Our models span a range of rotation periods and surface magnetic field strengths motivated by remnants of core-collapse supernovae, neutron-star mergers, and AIC of white dwarfs, and self-consistently incorporate neutrino emission and absorption during the early cooling phase.

Our principal finding is that sufficiently rapidly rotating ($P \sim 1$ ms), highly magnetized proto-magnetars ($B_{\rm p} \gtrsim 10^{15}$ G) naturally produce a strongly anisotropic outflow structure. Centrifugal support dramatically enhances mass loading at low latitudes, generating a dense, sub-relativistic equatorial wind, while polar field lines experience substantially weaker baryon loading. This angular stratification allows the polar outflow to attain magnetizations $\sigma \gtrsim 30$--$100$, despite neutrino luminosities of order $10^{52}\,\mathrm{erg\,s^{-1}}$, implying that ultra-relativistic motion is achievable on larger scales. In contrast to earlier angle-averaged or one-dimensional models, these results demonstrate that baryon loading does not preclude relativistic jet formation when multidimensional effects are taken into account.

A key result of our simulations is that jet collimation arises intrinsically at small radii ($\lesssim 100$ km), without requiring confinement by an external envelope of supernova or merger ejecta. The heavy equatorial wind acts as an effective confining medium, channeling magnetic flux and energy toward the rotation axis. This mechanism produces a narrow polar outflow whose energetics and opening angles are broadly consistent with those inferred for short gamma-ray bursts, while the majority of the wind power remains in slower, wider-angle ejecta that may instead contribute to the kinetic energy budget of supernovae or kilonovae. The resulting partition of energy between ultra-relativistic and sub-relativistic components provides a natural explanation for the coexistence of GRB emission and luminous, magnetar-powered transients in compact-object explosions.

Our results support a scenario in which a brief proto-magnetar phase can power a relativistic jet within seconds of remnant formation, potentially accounting for the prompt emission of short GRBs, while simultaneously driving more massive outflows that shape the longer-lived electromagnetic counterparts. The emergence of relativistic jets in this picture is therefore not an exceptional outcome, but rather a generic consequence of extreme rotation and magnetization during the earliest stages of PNS evolution. Future work will explore additional features of our magnetar wind models in detail and address their potential for heavy element nucleosynthesis (Desai et al., in prep).  

Several important extensions remain to be explored. First, our models initialize ordered, near-dipolar magnetic fields; future work should employ self-consistent remnant magnetic field configurations motivated by global dynamo processes seen in merger \citep{combi_jet_2023,kiuchi_dynamo_2024} and collapse simulations \citep{mosta_dynamo_2015,combi_aic_2025}, including non-axisymmetric and time-dependent field geometries. Second, the simulations presented here focus on quasi-steady wind states at fixed neutrino luminosity; following the full time-dependent evolution of proto-magnetar winds during cooling and spin-down will be essential for connecting jet properties to observed GRB durations and variability. Third, realistic merger \citep{combi_jet_2023} and rapidly rotating AIC \citep{micchi_multimessenger_2023,combi_aic_2025} remnants are surrounded by massive accretion disks, whose interaction with magnetar winds may significantly alter both collimation and baryon loading; incorporating disk–wind coupling in three-dimensional simulations is an important next step.  Fourth, as discussed in Sec.~\ref{sec:methods}, the use of more advanced neutrino transport schemes, such as a two-moment (M1) closure (e.g., \citealt{shibata_truncated_2011,foucart_monte_2018,li_neutrino_2021,radice2022new}) or Monte-Carlo scheme \citep{foucart_monte-carlo_2020}, would be desirable. 
Finally, extending these calculations to larger spatial scales will allow for a self-consistent study of how magnetar-driven jets and winds interact with kilonova or supernova ejecta, shaping observable light curves, spectra, and afterglows.

Our results strengthen the case for millisecond proto-magnetars as viable central engines of short gamma-ray bursts and related transients, and highlight the importance of multidimensional, neutrino-radiation GRMHD simulations in understanding the earliest moments of compact-object formation.

\begin{acknowledgments}
We thank Michael Müller and Aman Agarwal for helpful discussions. The authors gratefully acknowledge the computing time made available to them on the high-performance computers ``Emmy'' and ``Lise'' at the NHR Centers NHR@Göttingen and NHR@ZIB. These Centers are jointly supported by the Federal Ministry of Education and Research and the state governments
participating in the National High-Performance Computing (NHR) joint funding program (http://www.nhr-verein.de/en/our-partners). DD is supported by an Alexander von Humboldt Fellowship. DD and DMS acknowledge support by the Alexander von Humboldt foundation. This research was supported in part by Perimeter Institute for Theoretical Physics. Research at Perimeter Institute is supported in part by the Government of Canada through the Department of Innovation, Science and Economic Development Canada and by the Province of Ontario through the Ministry of Colleges and Universities. BDM acknowledges support from the National Science Foundation (grant AST-2406637) and the Simons Foundation (grant 727700). The Flatiron Institute is supported by the Simons Foundation. 

\end{acknowledgments}

\appendix

\section{Bernoulli Criterion}
\label{app:bern}

We discuss a few quantities for diagnostic purposes. The total specific energy that is conserved along the fluid path for an adiabatic stationary flow is given by the ratio between the energy flux and the rest-mass flux: 
\begin{equation}
    \mu = -T^{r}_t/\rho u^r \approx -(h+\sigma) u_t,
\end{equation}
where $u_t$ is the covariant time component of the 4-velocity, $h \equiv1+\epsilon+p/\rho$ is the relativistic specific enthalpy (with $\epsilon$ the specific internal energy). 
From this expression, we can define the Bernoulli parameter as $\mathcal{B} =\mu-h_{\infty}$, which is larger than zero ($\mu= h_{\infty} \Gamma_{\infty} > h_{\infty}$) if the fluid is unbound. The specific enthalpy $h_{\infty}$ at spatial infinity depends on the EOS through the definition of the zero point of $\epsilon$ ($h_{\infty}=1$ for our SFHo EOS table).
Unboundedness assumes that all thermal and magnetic energy is converted into kinetic energy at infinity (where spacetime is flat). 
The maximum Lorenz factor achievable by a fluid element is then determined as
\begin{equation}
    \Gamma_{\infty} = -u_t \frac{(h+\sigma)}{h_{\infty}}. \label{eq:gamma_inf}
\end{equation}
In this form, the Bernoulli parameter is also the total specific energy minus the rest-mass density, which we can separate into different components as:
\begin{equation}
    \mathcal{B} = \underbrace{(-u_t-1)}_{\rm kinetic} + \underbrace{-u_t (\epsilon + p/\rho)}_{\rm thermal} + \underbrace{- u_t \sigma}_{\rm magnetic}.
    \label{eq:bern}
\end{equation}
Note that it is only at large distances ($r\gtrsim 100$ km for our models) where $-u_t-1 \approx \Gamma-1$ is approximately the fluid specific kinetic energy.


\section{Analytic Estimates of Proto-Magnetar Outflow Magnetization}
\label{app:analytic}

\begin{figure}
    \centering
    \includegraphics[width=.5\linewidth]{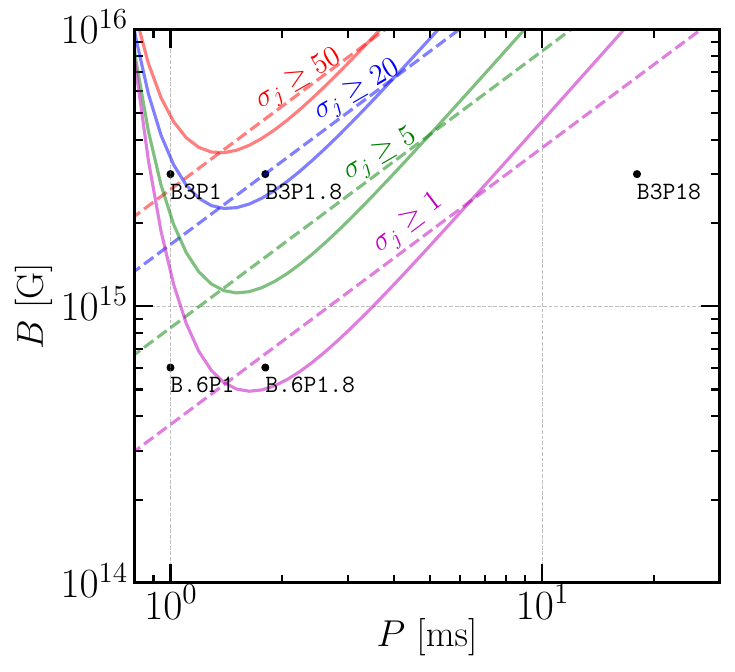}
    \vspace{-.3cm}
\caption{Contours of the magnetization of the equatorial wind ($\sigma_{\rm wind}$; solid lines; Eq.~\eqref{eq:sigma_wind}), and of the polar jet ($\sigma_{\rm j}$; dashed lines; Eq.~\eqref{eq:sigma_jet}) regions of the proto-magnetar outflow, in the parameter space of surface magnetic field strength $B_{\rm P}$ and spin period $P$, as estimated analytically in Appendix \ref{app:analytic} and calculated assuming $L_{\bar{\nu}_e} = 6\times 10^{51}$ erg s$^{-1}$ and $\theta_j =20^\circ$ for the size of the polar cap contributing to the polar jet.  The locations of proto-magnetar wind models are shown as solid points as marked.  Our most strongly magnetized and rapidly spinning models achieve $\sigma_{\rm j} \gtrsim 20-50$, showing that their jets may be capable of powering GRB emission.} \label{fig:sigma}
\end{figure}

We review results from the analytic model of proto-magnetar winds from \citet{metzger_proto-neutron_2007,Metzger+07b}, which accounts for rotational effects and the partial opening of field lines from a dipole to split monopole configuration. 

For unmagnetized, slowly-rotating spherical purely neutrino-driven outflows, the total wind mass-loss rate can be approximately written as (e.g., \citealt{Qian&Woosley96})
\be
\dot M_{\rm sph} \approx 3\E{-5} L^{5/3}_{52} \epsilon^{10/3}_{18} R_{10}^{5/3} M^{-2}_{1.4} ~M_\odot\,{\rm s^{-1}},
\label{eq:Mdotsph}
\ee
where $L_{52} \equiv L_{\bar \nu_e}/10^{52}\mathrm{erg~s}^{-1}$, $\epsilon_{18} \equiv \epsilon_{\bar \nu_e}/ 18 \mathrm{~MeV}$, $R_{10} \equiv R_{\bar \nu_e}/ 10 \mathrm{~km}$, and  $M_{1.4} \equiv M/ 1.4 M_\odot$. This estimate accounts for the net neutrino heating in the quasi-hydrostatic gain region, and the prefactor has been calibrated to match the (roughly isotropic) mass-loss rate found in the \texttt{NRNM} model (Fig.~\ref{fig:angular}).

Eq.~\eqref{eq:Mdotsph} assumes an outflow can occur from the entire magnetar surface.  However, in the presence of a strong magnetic field, magnetic tension blocks such an outflow across a portion of the surface, i.e. $\dot{M} \rightarrow f_{\rm open}\dot M_{\rm sph}$.  Here, $f_{\rm open} = 1-\cos \theta_{\rm LCFL} < 1$ is the fraction of the surface area threaded by open magnetic field lines, where $\theta_{\rm LCFL}$ is the polar angle of the last closed field line (i.e., that which returns to the magnetar surface instead of feeding the unbound outflow).  Approximating the magnetosphere as a dipole field, the last closed field line intersects the equator at the $Y$-point radius, $R_Y$, such that $\theta_{\rm LCFL} \simeq \arcsin\sqrt{R_{\nu}/R_Y}$.   

In force-free pulsar winds, one typically expects $R_{\rm Y} \approx R_{\rm L}$, where $R_{\rm L} = c/\Omega = 2\pi P c$ is the light-cylinder radius, and the resulting open magnetic flux gives rise to the standard dipole spin-down energy loss rate (e.g., \citealt{Spitkovsky06}).  However, in general $R_{\rm Y} < R_{\rm L}$ as a result of the additional inertia from mass-loading of the field lines (e.g., \citealt{Mestel&Spruit87, Bucciantini+06}), leading to enhanced spin-down and a greater total mass-loss rate.  A pure split monopole configuration corresponds to the limit that $R_{\rm Y} \rightarrow R_{\nu}$ and $f_{\rm open} \rightarrow 1$. 

On the one hand, strong magnetic fields inhibit mass-loss by closing off a portion of the magnetar surface to outflows.  On the other hand, they also cause the atmosphere to co-rotate with the star close to its surface, which reduces the effective gravity and expands the pressure scale-height in the gain region.  For those regions of the surface open to unbound outflows, this acts to boost the mass-loss rate per unit surface area (e.g., \citealt{metzger_proto-neutron_2007}), exponentially in proportion to the surface rotational speed near the last-closed field line, $v^\phi_{\rm LCFL} = 2\pi R_\nu \sin \theta_{\rm LCFL}/P$ (since field lines with $\theta \lesssim \theta_{\rm LCFL}$ dominate the total open surface area).  Following \citet{Metzger+07b}, the wind mass-loss rate is boosted from the non-rotating value (Eq.~\eqref{eq:Mdotsph}) approximately according to,  
\begin{eqnarray}
\dot M_{\rm cf} \approx f_{\rm open}\dot{M}_{\rm sph} \exp \left(\frac{v^\phi_{\rm LCFL}}{c_s}\right)^{2} \approx
f_{\rm open}\dot{M}_{\rm sph} 
\exp \left( 4~ P_{\rm ms}^{-2} R_{10}^2 \frac{R_\nu}{R_Y}c_{s,0.1}^{-2} \right),  
\label{eq:Mdot_boost}
\end{eqnarray}
where $c_s\approx 0.1 c~ L^{1/12}_{52} \epsilon^{1/6}_{18} R_{10}^{-1/6}$ is the surface sound speed, $c_{s,0.1} \equiv c_s/0.1c$, and $P_{\rm ms} \equiv P/1~\rm ms$.  

The total open magnetic flux threading the outflow can be estimated as $\Phi_{\rm B} = \int B \cdot dS \approx (f_{\rm open}\pi R_\nu^2)B_{\rm P},$ where $B_{\rm P}$ is the magnetic field strength near the polar cap (the magnetic field of a dipole only changes by a factor of 2 from the pole to the equator).  The outflow magnetization $\sigma$, also representing the ratio of Poynting flux $\propto \Phi_{B}^{2}$ to mass flux $\dot{M}$, can be written (e.g., \citealt{metzger_magnetar_2018}):
\begin{eqnarray}
\sigma_{\rm wind} &\equiv& \frac{\Phi_{B}^{2}\Omega^{2}}{\dot{M}c^{3}} =\frac{ \pi^2 B_{\rm P}^{2}R_{\nu}^4 f^2_{\rm open}\Omega^{2}}{\dot{M}_{\rm cf}c^{3}}  
\approx  1~R_{10}^{7/3}  B_{15}^2  P_{\rm ms}^{-2} L^{-5/3}_{52} \epsilon^{-10/3}_{18} M^{2}_{1.4} f_{\rm open} 
\exp \left( -4~ P_{\rm ms}^{-2} R_{10}^2 \frac{R_\nu}{R_Y}c_{s,0.1}^{-2} \right) 
\label{eq:sigma_wind},
\end{eqnarray}
where $B_{15} = B_{\rm P}/(10^{15}$ G$)$.  Note the dual role played by $f_{\rm open}$ in setting both the open magnetic flux and mass-loss rate of the wind.  We use the value of $R_{\bar \nu_e}/R_Y$ measured in our simulation to estimate $f_{\rm open} \approx 1- \sqrt{1-R_\nu/R_Y}$.  

Eq.~\eqref{eq:sigma_wind} reasonably approximates the luminosity-weighted average magnetization of the magnetar wind, as the bulk of the outflow occurs along field lines with $\theta \sim \theta_{\rm LCFL}$ which carry most of the wind energy (e.g., \citealt{Metzger+11a}), consistent with the minimum value of $\sigma$ achieved within the equatorial outflows in our simulations (Fig.~\ref{fig:angular}).  

However, because of the exponential sensitivity of the wind mass-loss rate to the polar angle for millisecond spin periods (Eq.~\eqref{eq:Mdot_boost}), Eq.~\eqref{eq:sigma_wind} will under-estimate the magnetization of the polar outflow whichs end up threading the ``jet''.  We define the jet as those field lines which leave the magnetar surface up to the critical polar angle, $\theta_{\rm j}$, for which mass-loading is not significantly enhanced by centrifugal effects and hence $\sigma \simeq \sigma_{\rm j}$ reaches its maximum value. More quantitatively, $\theta_j$ can be defined as the polar angle above which the rotational velocity, $v^\phi = R_\nu\Omega \sin \theta_j$ exceeds the surface sound speed $c_s$.  For outflows which leave the PNS close to the equator, centrifugal effects become important below a critical rotation period:
\be 
P_{\rm cf} \simeq 2\pi R_{\nu}/c_{s} \simeq 2 ~L^{-1/12}_{52} \epsilon^{-1/6}_{18} R_{10}^{7/6} \mathrm{~ms}. \label{eq:Pcf}
\ee

Replacing $f_{\rm open} \rightarrow \theta_{\rm j}^{2}/2$ and $\dot{M} \rightarrow f_{\rm open}\dot{M}_{\rm sph}$, Eq.~\eqref{eq:sigma_wind} can be transformed into an estimate of the jet magnetization,
\begin{align}
\sigma_j \equiv \frac{ \pi^2 B_{\rm P}^{2}R_{\nu}^4f^2_{\rm open}\Omega^{2}}{f_{\rm open}\dot{M}_{\rm sph}c^{3}}   \approx  4.5 ~ R_{10}^{7/3}  B_{15}^2  P_{\rm ms}^{-2}  \left(\frac{\theta_{\rm j}}{20^\circ}\right)^2 L^{-5/3}_{52} \epsilon^{-10/3}_{18} M^2_{1.4}.
\label{eq:sigma_jet}
\end{align}
We compare the maximum magnetization of the polar jet found in our models to this expression in Fig.~\ref{fig:sigma}.

\section{Spatial resolution and convergence}
\label{app:resolution}

\begin{figure}
    \centering
\includegraphics[trim={0cm 0cm .2cm 0cm},clip,width=.56\linewidth]{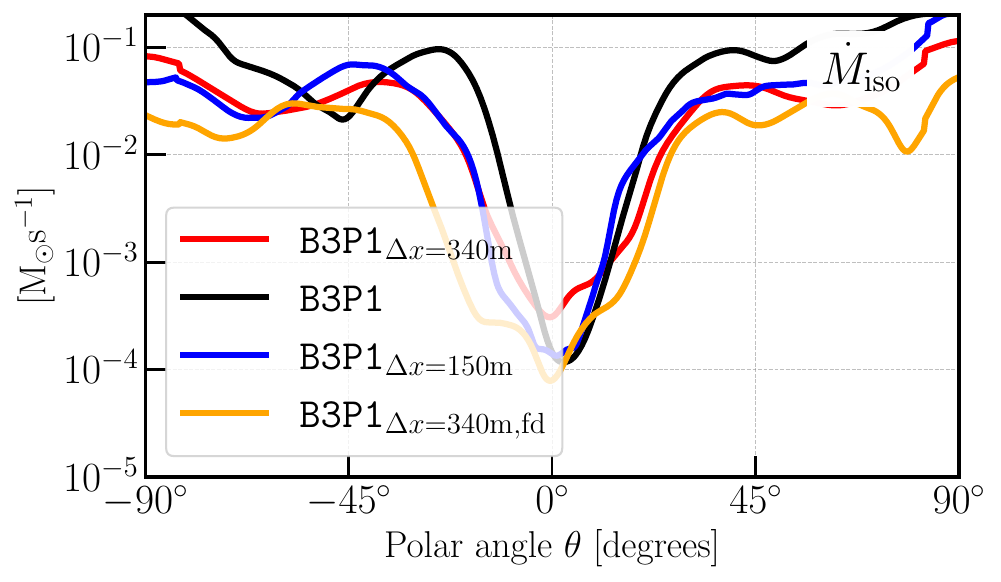}\includegraphics[trim={3cm -.3cm .3cm 0cm},clip,width=.55\linewidth]{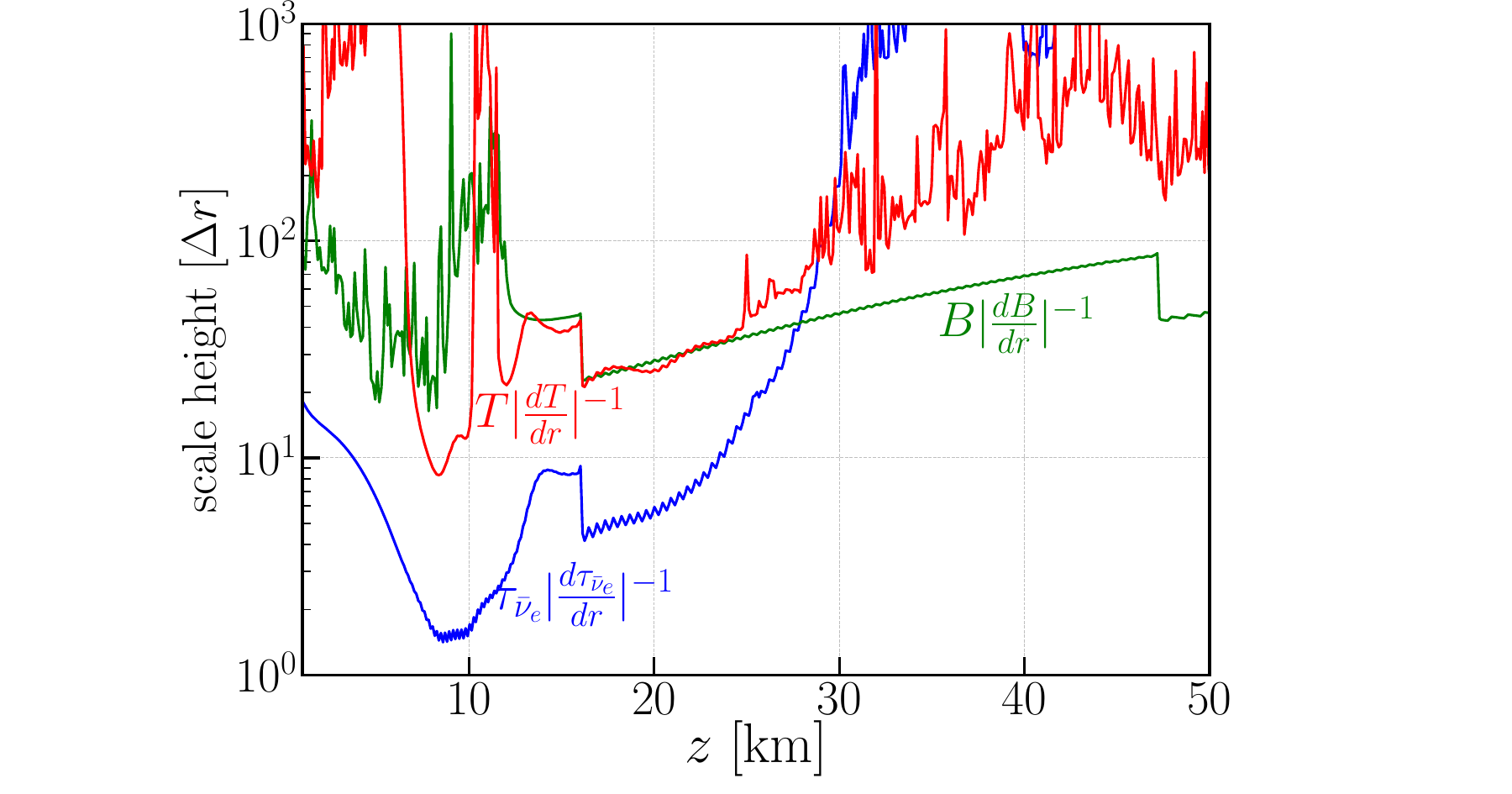}
\vspace{-.7cm}
\caption{\textit{Left:} Time-averaged angular profiles of isotropic-equivalent mass-loss rate $\dot M_{\rm iso}$ through a shell of radius $R=120$ km for model \texttt{B3P1} at three different grid resolutions with smallest grid step size $\Delta x=340$ m (red), $\Delta x=225$ m (fiducial resolution, black), $\Delta x=150$ m (blue), as well as one from a simulation without using symmetries (labeled ``full domain -- fd'') and with resolution $\Delta x=340$ m (orange). \textit{Right}: Scale heights of temperature ($T |\frac{dT}{dr}|^{-1}$), optical depth of electron anti-neutrinos ($\tau_{\bar \nu_e} |\frac{d\tau_{\bar \nu_e}}{dr}|^{-1}$) and magnetic field ($B|\frac{dB}{dr}|^{-1}$) in the polar direction in units of the grid resolution plotted as a function of the $z$-coordinate, cylindrically-averaged within a cylindrical radius of 5 km from 2D cross-sectional data of the $xz$-plane for model \texttt{B3P1}.} \label{fig:convergence}
\end{figure}

The left panel of Fig.~\ref{fig:convergence} shows results from a convergence test of our simulation suite. The outflow structure as manifested in the isotropic-equivalent mass-loss rate across three different resolutions with finest spacings $\Delta x=340$\,m, $\Delta x=225$\,m (fiducial resolution), and $\Delta x=150$\,m is nearly identical, considering that spatio-temporal fluctuations are inherent to the flow structures, which complicates direct comparison. We show additionally the outflow structure of the mass-loss rate for outflows using no symmetries (labeled ``$\Delta x = 340$,fd''), and this outflow structure is also nearly identical to the original model. Furthermore, the full-domain, high, and low-resolution simulations could not be run over long physical time scales due to computational constraints, such that flow structures are not perfectly stationary out to large radii, further complicating direct comparison. For instance, the full-domain simulation has a slightly larger opening angle of the jet structure, because it is still approaching steady-state. 

As an additional check on our fiducial resolution, the right panel of Fig.~\ref{fig:convergence} shows that all necessary scale heights are resolved throughout nearly the entire radial domain with at least $\gtrsim\!10$ grid points per scale height, sufficient to properly resolve the dynamics. The exception to this is the neutrino scale height in the neutrinosphere region. As discussed in Papers I and II, this may result in inaccuracies of neutrino energies and luminosities; for our purposes, however, we do not wish to pinpoint the precise neutrino energies/luminosities. Rather, we treat these as parameters during the proto-magnetar cooling evolution. Their precise values are sensitive to initial conditions prior to proto-magnetar formation (e.g., as originating in a core-collapse scenario). Neutrino heating, on the other hand, occurs in a ‘gain region’ $\sim$tens of kilometers above the proto-magnetar surface, which is well resolved with $\gtrsim 10$ grid points per scale height.


\bibliographystyle{aasjournalv7}

\bibliography{mainrefsPNS,Relastro_PI_UoG}

\end{document}